\documentstyle[aps, floats, epsf, twocolumn]{revtex}

\begin{document}

\title{$QED_3$ theory of underdoped high temperature superconductors}

\author{Igor F. Herbut}

\address{Department of Physics, Simon Fraser University, 
Burnaby, British Columbia, \\
 Canada V5A 1S6\\} \maketitle

\begin{abstract}

The low-energy theory of d-wave quasiparticles 
coupled to fluctuating vortex loops that describes the loss of 
phase coherence in a two dimensional d-wave
superconductor at $T=0$ is derived from first principles.  The theory
has the form of 2+1 dimensional quantum electrodynamics ($QED_3$),
and is proposed as an effective description of the $T=0$
superconductor-insulator transition and of the pseudogap phase
in underdoped cuprates. The coupling constant ("charge")
in this theory is proportional to the dual order parameter of the
XY model, which is assumed to be describing fluctuations
of the phase of the superconducting order parameter.
Finiteness of the charge is then tantamount to the appearance of
infinitely large vortex loops, i. e. to the loss of phase coherence
in the system. The principal result is that the destruction of 
superconducting phase coherence in the d-wave superconductors typically, and 
immediately, leads to the appearance of antiferromagnetism.
This transition can be understood in terms of the spontaneous 
breaking of an approximate "chiral" $SU_c (2)$
 symmetry, which may be discerned at low enough energies in the standard 
d-wave superconductor. The mechanism of this spontaneous
symmetry breaking is formally analogous to the
dynamical mass generation in the $QED_3$, with the "mass" here being
proportional to staggered magnetization. Other phases with broken chiral
symmetry include the translationally invariant "d+ip" and
"d+is" insulators, and
the one-dimensional charge-density and spin-density waves, which are all 
 insulating descendants of the d-wave superconductor. 
All the insulating states have the neutral
spin-$1/2$ excitations that one can identify in the superconductor
confined by the logarithmic potential. Electron repulsion 
is in this formalism
represented by a particular quartic perturbation to the $QED_3$ action,
which breaks the chiral symmetry and selects the 
antiferromagnet as the preferred broken symmetry state. I formulate the
mean-field theory of the antiferromagnetic instability in presence
of a short-range repulsive interaction, and find the
staggered magnetization to be significantly enhanced
deeper inside the insulating state.  The theory offers an
explanation for the rounded d-wave-like dispersion seen in ARPES
experiments on the  insulating $Ca_2 Cu O_2 Cl_2$
(F. Ronning {\it et. al.}, Science {\bf 282}, 2067 (1998).) 
Relations to other theoretical approaches to
 the high-$T_c$ problem are discussed.
\end{abstract}

PACS: 74.20.Mn, 74.25.Jb, 74.40.+k

\vspace{10pt}
\section{Introduction}

Soon after the original discovery,
it became  well appreciated that the high temperature (high-$T_c$) 
superconductors are  all
quasi two-dimensional insulating antiferromagnets  that become
superconducting with the introduction of holes.
The nature of the relationship between antiferromagnetism and
high temperature superconductivity has been the central issue in 
the field. Following the time honored strategy of understanding first 
the non-superconducting state, most of the approaches to the 
high-$T_c$ problem focused on finding the mechanism by which doping an  
antiferromagnet would produce a superconductor \cite{anderson}.
The essential difficulty in pursuing this  
strategy seems to be that the Mott insulator is itself a non-trivial 
strongly correlated state, harder to describe in simple terms than
the metallic Fermi liquid, which played its role in the BCS theory of the
low-temperature superconductivity \cite{schrieffer}. The situation
becomes only worse away from half-filling, where the ground state
of even the simplest models becomes more ambiguous. Experimentally,
the cuprates seem to loose their antiferromagnetic ordering with
doping before they
become superconducting, and many candidates for the
intermediate "pseudogap phase" have been discussed in literature. 
The nature of the non-superconducting
state that is supposed to be unstable to superconductivity with doping
is at this point, however, far from clear, and may prove to be
non-universal. Arguably, the physics of underdoped regime may be 
the main mystery of high temperature superconductivity. 

 In a remarkable contrast to the uncertainties inherent to the 
insulating phase, the superconducting phase of most high-$T_c$ materials
is well established to have  the d-wave symmetry of the
order parameter \cite{hardy}, \cite{wollman}, typically 
with well-defined, long-lived  quasiparticle
excitations \cite{hosseini}, \cite{feng}.
This simplicity suggests that an {\it inverted} approach to the high-$T_c$
problem may be more natural \cite{balents}:
if there exists a d-wave state in the phase diagram,
which {\it other} states can in principle be inferred from it?
The purpose of this paper is to establish the theoretical framework for
answering this question, answer it,
and show how this may help explain some salient features of the cuprates
phase diagram and the angle resolved photo emission spectroscopy (ARPES)
experiments in the insulating state \cite{ronning}, \cite{ronning1}.

 Loosely speaking, there are two ways to destroy a superconducting state:
1) by driving the amplitude of the order parameter to zero, which
is what is well described by the weak-coupling BCS theory
at finite temperature \cite{schrieffer}, for example. For a d-wave
superconductors this process presumably is relevant at large dopings,
where weak-coupling treatments of the Hubbard and related models
 can be trusted,
and disorder should eventually force $T_c$ to vanish \cite{herbut1}.
2) Even if the amplitude of the order parameter
is large and finite, superconductivity will be lost with the
destruction of phase order \cite{doniach}, \cite{kivelson}.
There is evidence
that this is what actually occurs in underdoped cuprates, where the
superconducting transition temperature ($T_c$)
 is much lower that the pseudogap temperature $T^*$.
Since underdoped cuprates are strongly two dimensional, at finite
temperatures the loss of phase order may be expected to proceed via 
the Kosterlitz-Thouless transition, and indeed, there are  distinct 
experimental signatures of the fluctuating vortices above
$T_c$ \cite{corson}, \cite{xu}.
The following question then naturally arises:
What is the nature of the $T=0$ phase that derives from a two-dimensional
d-wave superconductor when the phase-coherence is lost, but the order
parameter amplitude is still finite? The central thesis of this work
is that phase incoherent d-wave superconductor (dSC) is nothing but
the insulating (typically incommensurate) spin-density-wave (SDW), i. e.
weak antiferromagnet. Short account of this result appeared earlier
in \cite{herbut2}.

I show that the minimal continuum theory of the low-energy quasiparticle
excitations near the four nodes of the d-wave order parameter
coupled to fluctuating vortex loops at $T=0$
is provided by the $2+1$ dimensional
quantum electrodynamics ($QED_3$):
\begin{equation}
S=  \int d^2 \vec{r} d\tau [  \bar{\Psi}_i  \gamma_\nu
(\partial_\nu + i a_\nu) \Psi_i  + \frac{1}{2|\langle \Phi \rangle|^2}
(\nabla\times \vec{a})^2 ], 
\end{equation}
where $\nu=0$, (imaginary time) $1,2$ (space),
and the sum over repeated indices is assumed.
The {\it four-component} Dirac fields $\Psi_i $ $i=1,2$ represent the sharp,
electrically neutral spin-1/2 excitations one can define
in the superconducting state (and hence may call "spinons"),
which are minimally coupled to a massless  
gauge-field $\vec{a}$. The gauge-field 
derives from the fluctuating topological defects (vortex loops) in the
phase of the superconducting order parameter, which have been
integrated out in deriving the theory (1).
Complex number $\langle \Phi \rangle $ is proportional to the
the disorder (dual order) parameter \cite{kleinert0}, and
represents the state of vortex loops: $\langle \Phi \rangle
\neq 0$ signals the appearance of infinitely large loops in the system
and the loss of phase coherence,
which is the $T=0$ analog of the Kosterlitz-Thouless transition \cite{nguyen}.
In the superconducting state, on the other hand, all  
 loops are of finite size, $\langle\Phi\rangle=0$, and the
gauge field, in the simplest approximation, may be
considered effectively decoupled from the spinons: quasiparticle excitations
are then sharp, since all the short-range
interactions that have not been explicitly written in the Eq. (1),
if weak enough, are strongly irrelevant. When $\langle \Phi \rangle \neq 0$
the situation becomes radically different, as the gauge-field  
mediates a long-range interaction between spinons. In reality
the theory is also strongly anisotropic, but for simplicity this
possibly important feature
has been neglected in writing the Eq. (1). The $QED_3$ has also been
recently considered by Franz and Te\v sanovi\' c \cite{franz}
 as an effective description  of the pseudogap state.
They argued that the presence of the massless
gauge field may explain the broad features seen in ARPES
measurements in the normal state \cite{feng}, \cite{valla}.
Here I show that at $T=0$ as soon
as $\langle \Phi \rangle$ becomes finite there is a dynamical  
generation of the mass term $\sim m\bar{\Psi}_i \Psi_i $ in (1),
which can be identified as the staggered potential felt by the
original electrons, i. e. with the SDW order parameter.
Quantum fluctuating dSC is thus at $T=0$ inherently unstable towards
SDW ordering once the phase coherence is lost.  

\begin{figure}
\centerline{\epsfxsize=7cm \epsfbox{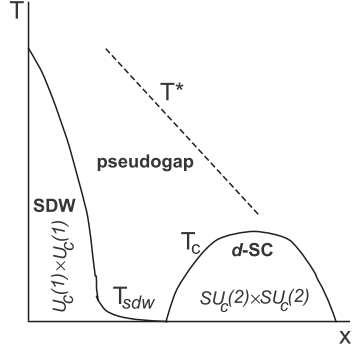}}
\vspace*{1em}

\noindent
\caption{ A schematic phase diagram of cuprate superconductors in terms
of the low-energy chiral symmetries $SU_c (2)$ (full) and
$U_c (1)$ (broken). Besides the chiral symmetries, d-wave superconductor
(dSC) also has the full spin rotational symmetry, and the spin-density wave
(SDW) has the superconducting $U(1)$ and the spin rotational symmetry around
one of the axis. Near and left to the underdoped
transition point the system is proposed to be an extremely weak SDW,
which becomes reinforced by the electron repulsion,  
and which continuously evolves into a stronger antiferromagnet near
half-filling.}
\label{f1}
\end{figure}

   The dSC $\rightarrow$ SDW quantum phase transition is an
example of spontaneous breaking of a continuous
global symmetry in the Eq. (1), which for a lack of better name I will
call "chiral" throughout the paper. Chiral symmetry breaking is a well
studied field-theoretic phenomenon, believed to be
inextricably linked to  confinement in the $QED_3$ \cite{appelquist}. 
Massless $QED_3$ for single species of Dirac fermions
has the continuous $U(2)=U(1) \times SU_c (2)$ symmetry, with the generators
$I,\gamma_3, \gamma_5$, and $\gamma_{35}= i\gamma_3 \gamma_5$, respectively.
In the action in the Eq. (1), the $U(1)$ factor represents the residual
spin rotational symmetry left by the choice of representation, as
will be explained in
detail later. It is the additional $SU_c (2)$ symmetry per Dirac component in the
$QED_3$ that will be of central interest here. The fermion
mass term $m\bar{\Psi}_i \bar{\Psi}$ breaks the $SU_c (2)$ for each Dirac
field to $U_c (1)$, and the two broken generators
rotate between different  insulating states.
Chiral $SU_c (2)$ arises as an approximate symmetry
of the dSC only at low-energies, and will be
manifestly broken, for example, by higher order
derivatives omitted in the Eq. (1). It
should not be confused with the spin rotational symmetry
which is, of course, also, and exactly,  present in the dSC. Higher order 
derivatives and the electron interaction terms reduce the $SU_c (2)$ to
its $U_c (1)$ subgroup, which is related to the 
spatial translations of the original electrons.
The identification of the approximate  chiral symmetry in the
dSC is essential for establishing the connection
between the antiferromagnetic and the superconducting phases
advocated in this paper, and represents one of my central results.
The idealized cuprates phase diagram may be understood in terms of the
chiral symmetries of different states as depicted in Fig. 1.

Assuming the scale for the SDW transition  $T_{SDW} (x) $ in an
anisotropic quasi-two dimensional high-temperature superconductor to be
set by the magnitude of the staggered magnetization
at $T=0$ \cite{keimer},  the present work suggests that 
near and left of the superconductor-insulator transition one should
expect it to be considerably lower than the 
superconducting $T_c (x)$ near and right of the critical point:
$T_{SDW}(x_u - \delta) << T_c (x_u +\delta) $, where
$x_u$ is the critical doping for the dSC-SDW transition, and $\delta <<1$
(see Fig. 1).
This is because the generalized $QED_3$ with $N$ fermion species has
a critical point at $N=N_c \approx 3$, above which
there is no dynamical mass generation \cite{appelquist}.
The $QED_3$ in (1) has $N=2$ components, which together with 
some numerical factors gives very weak SDW order near the
superconducting phase. The pseudogap
phase in cuprates at $T=0$ is therefore proposed here to be actually 
an extremely fragile SDW, likely to be  easily destroyed by disorder, for
example. As half-filling is approached and the vortex loop
condensate $\langle \Phi \rangle $ increases, the repulsion between electrons
also becomes important. Short-range repulsion  is
represented in the $QED_3$ by a particular quartic term, which if weak 
is irrelevant in the superconducting state, but which also manifestly 
breaks the chiral symmetry of the low-energy theory. I
show that the effect of such a term
is first to break the degeneracy among states with broken chiral symmetry
in favor of the SDW, and then to 
dramatically increase the SDW order parameter farther from the dSC.
The picture implied by the $ QED_3$
is qualitatively in accord with the generic phase diagram for
the underdoped cuprates, where the antiferromagnetic transition 
near half-filling raises to $\sim 300 K$, but is typically
unobservably low very near the superconducting state.

Neutral spinons, which are well
defined quasiparticles in the superconducting state, in the insulator
become broad excitations with the lifetime proportional to the
antiferromagnetic order parameter. At $T=0$
and at large distances they  become confined by a logarithmic potential
provided by the gauge-field in presence of the chiral symmetry
breaking. Due to the weakness of the SDW order very
near the superconducting transition,
however, spinon confinement is effective only
at very large distances, or equivalently, at very low temperatures.
The weak SDW phase therefore appears effectively
deconfined at intermediate length scales. The finite-T pseudogap
phase has the gapless spinons strongly scattered by the
massless gauge field, in qualitative agreement with the broad
spectral features of the electrons seen in ARPES \cite{franz}.
Near half-filling the SDW order increases and the bound state
of spinons rapidly shrinks, leaving only magnons in the excitation spectrum.

Confined nature of the standard antiferromagnet close to half-filling,
if postulated, 
by itself already points to the $QED_3$ as a viable candidate for the 
effective theory of underdoped cuprates.
If one views the superconducting state
as being spin-charge separated \cite{balents}, one needs a
mechanism by which spinons would eventually become confined in the
antiferromagnetic phase. The $QED_3$ provides such a mechanism
automatically, since the massless gauge-field mediates a long-range
logarithmic interaction between the spinons that binds them
at all energies. Were the gauge-field massive, on the
other hand, the physics would be equivalent to $Z_2$ gauge theory,
and the antiferromagnetic state would be
deconfined and quite different from the usual antiferromagnet
\cite{wen}, \cite{mudry}, 
\cite{senthil}. The very existence of an ordinary antiferromagnet at, and
presumably near, half filling \cite{kastner} may
therefore be taken as evidence in favor of the type of theory presented
in this paper.

The physical picture of the antiferromagnetic (SDW) insulator as a 
phase-disordered d-wave superconductor is further
 supported by the ARPES data
on the insulating $Ca_2 Cu O_2 Cl_2$ and  $Sr_2 Cu O_2 Cl_2$
\cite{ronning}, \cite{ronning1}. These experiments show two
unexpected features of the insulating state: 1) although
the ARPES spectral function is broad, one can nevertheless identify
a remnant of the Fermi surface, 2) the 
dispersion at such an approximate "Fermi surface"  has a d-wave
form, except that it becomes rounded and without the characteristic cusp
at low energies. The "relativistic" dispersion for broad 
quasiparticle excitations that the $QED_3$  implies
in the insulating state, when measured from the lowest energy
given by the dynamically generated chiral mass, provides a very good fit to
the data (see Fig. 5). The present theory implies that the rounding
of the dispersion is controlled by the size of the sublattice magnetization,
and therefore should decrease with doping, as one approaches the
superconducting state. It would be desirable to test this prediction in
future experiments.

 In the body of the paper I develop the above picture in detail.
 In the next section,
 I derive the Dirac representation of the Hamiltonian for low-energy nodal
 quasiparticles, and discuss the
 coupling to quantum fluctuating vortex loops in the
 section III. A derivation of
 the dynamics of the gauge field starting from the XY
 model on a lattice is presented in the sec. IV. This section is somewhat 
 technical and may be skipped at first reading. Instead, the
 reader may consult the Appendix B, where a simpler derivation
 for finite temperatures is presented. Dynamical breaking
 of the chiral symmetry and the formation of the SDW state
 is discussed in the sec. V. More general discussion of the chiral
 symmetry and the other ordered states
 on the chiral manifold is provided in the sec. VI. The reduction 
 of chiral symmetry by the irrelevant terms
 is discussed in sections VII, and the mean-field theory of the
 antiferromagnetic instability of the $QED_3$ in presence of the
 electron repulsion is solved
 in the sec. VIII. Confinement of spinons in the insulator
 is discussed in the sec. IX. The 
 discussion of the ARPES  measurements is given in the sec. X. 
 Summary of the main results and the discussion of the relations to 
 other theoretical approaches is given in the concluding
 section. I finish with the list of open problems.
 Technical details are presented in five Appendices.

\section{Dirac theory for nodal excitations}

I begin by assuming that the superconducting state,
except from  being a d-wave, 
otherwise exhibits the standard BCS phenomenology. In particular,
I take that the quasiparticles are well-defined,
long-lived excitations.  Generally, the
quasiparticle action  at $T\neq 0$ may then be taken to be
\begin{eqnarray}
S = T \sum _{\vec{k}, \sigma, \omega_n }[  (i\omega_n -\xi_{\vec{k}
}) c^{\dagger}
_{\sigma}(\vec{k},\omega_n ) c_{\sigma} (\vec{k},\omega_n ) \\ \nonumber 
-  \frac{\sigma}{2}     \Delta(\vec{k})  c^{\dagger}
_{\sigma}(\vec{k},\omega_n ) c^{\dagger}_{-\sigma} (-\vec{k},- \omega_n )
+ h. c.  +O( c^4 )],
\end{eqnarray}                                                       
where $\Delta(\vec{k})$ has the usual d-wave symmetry, and 
two spatial dimensions (2D) are assumed. $c$ and $c^{\dagger}$ are the
electron operators, $\sigma=\pm$ labels the z-projection of
electron spin, and $\omega_n $ are
the fermionic Matsubara frequencies. Units are chosen so that
$h=c=e=1$. $O(c^4)$ term stands for all possible short-range
interactions between quasiparticles.

We may represent the quasiparticle Hamiltonian in terms of
two {\it four-component} fields,
\begin{eqnarray}
\Psi^{' \dagger} _{i} (\vec{q},\omega_n ) = (c^{\dagger}_+ (\vec{k},\omega_n ),
c_- (-\vec{k}, -\omega_n ), \\ \nonumber
c^{\dagger}_+ (\vec{k}-\vec{Q}_{i}, \omega_n ),
c_- (-\vec{k}+\vec{Q}_{i}, -\omega_n ) ), 
\end{eqnarray}
where $\vec{Q}_{i} = 2\vec{K}_{i}$ is the wave vector that connects
the nodes within the diagonal pair $i=1,2$, as in Fig. 2.
For spinor 1, $\vec{k}=\vec{K}_1 +\vec{q}$, with $|\vec{q}|\ll |\vec{K_1}|$,
and analogously for the second pair. The construction of the four-component
field is not unique. The choice in the Eq. (3)
differs from the one made in the ref. \cite{balents}, for example.
I postpone the discussion of  the alternative construction used there 
for the Appendix D.
Using the construction in the Eq. (3), and by observing  that   
$\xi_{\vec{k}}= - \xi_{ \vec{k} - \vec{Q}_{i} }$, and 
$\Delta_{\vec{k}}= - \Delta_{ \vec{k} - \vec{Q}_{i} }$,
for $\vec{k}\approx \vec{K}_{i}$, and then by linearizing  the
spectrum as $\xi_{\vec{k}} = v_f q_x+ O(q^2)$
and $\Delta_{\vec{k}} = v_\Delta q_y +O(q^2) $,
one arrives at the low-energy action
\begin{eqnarray}
S[\Psi ' ] = \int d^2 \vec{r} \int_0 ^ {\beta} d\tau \Psi_1  ^{' \dagger} [
\partial_\tau
+ M_1 v_f \partial_x + M_2 v_{\Delta} \partial_y] \Psi_1 ' \\ \nonumber
 + ( 1 \rightarrow 2, x\leftrightarrow  y) + O(\partial \Psi ^{' \dagger}
  \partial \Psi ' , \Psi ^{' 4}), 
\end{eqnarray}
with $\beta = 1/T$. 
The continuous Dirac field $\Psi' _i (\vec{r},\tau)$ is defined as 
\begin{eqnarray}
\Psi_{i} ' (\vec{r},\tau) = T\sum_{\omega_n} \int \frac{d^2 \vec{q}}
{ (2\pi)^2 } e^ {i\omega_n \tau
+i \vec{q}\cdot \vec{r}} \Psi_{i} '  ( \vec{q}, \omega_n), 
\end{eqnarray}
with the integral over momenta performed over $|\vec{q}|< \Lambda < 
T^*$. The $4\times 4$ matrices in the Eq. (4)
are $M_1 = i \sigma_3 \otimes \sigma_3 $, and
$M_2 = -i \sigma_3 \otimes \sigma_1$. $\vec{\sigma}$ are the
usual Pauli matrices, and the coordinate system has been rotated
as in Fig. 2.

To cast the theory in Dirac form we may invoke the matrix
$\gamma_0= \sigma_1 \otimes I$, where $I$ is the $2\times 2$ unit
matrix. Then $\gamma_0 ^2 =I\otimes I$, and
$M_{i} = \gamma_0 \gamma_{i}$, with $\gamma_1 = \sigma_2 \otimes
\sigma_3$, and $\gamma_2 = - \sigma_2 \otimes \sigma_1$.
$\{ \gamma_\nu, \gamma_\mu \} = 2 \delta_{\nu \mu}$,
$\nu, \mu =0,1,2$, so the $\gamma$-matrices
indeed satisfy the Clifford algebra.
The quasiparticle action  (2) at {\it low energies}
becomes equivalent to the field theory
\begin{eqnarray}
S[\Psi' ] = \int d^2 \vec{r} \int_0 ^ {\beta} d\tau \bar{\Psi} _1 ' [
\gamma_0 \partial_\tau
+ \gamma_1 v_f \partial_x + \gamma_2 v_{\Delta} \partial_y] \Psi_1 '
\\ \nonumber
 + ( 1 \rightarrow 2, x\leftrightarrow y ) +O(\partial \bar{\Psi}' 
  \partial \Psi ' , \Psi ^{' 4} ) ,
\end{eqnarray}
where $\bar{\Psi}' _i =\Psi_i ^{' \dagger} \gamma_0$. Weak quartic interactions, as
long as they are short-ranged, are irrelevant by simple power counting.
This simply reflects the severe phase-space restrictions for
scattering of the nodal quasiparticles. I will therefore omit them
temporarily, together with the second order derivative
terms, ato return to their effects in the section VII.

The reader would be correct to
note that there is a considerable freedom in selecting
the form of the matrix $\gamma_0$. In fact, any $4\times 4$
matrix that anticommutes with
$M_1$ and $M_2$ and squares to unit matrix
 would yield an equally valid Dirac
representation. It is shown later that this freedom will correspond to
different "directions" in the space of ordered states with broken
chiral symmetry. The specific choice for $\gamma_0$ 
made here will be 
 analogous to choosing a direction in real space along which
to search for a finite magnetization, for example, in the more 
familiar magnetic phase transitions.

\begin{figure}
\centerline{\epsfxsize=5cm \epsfbox{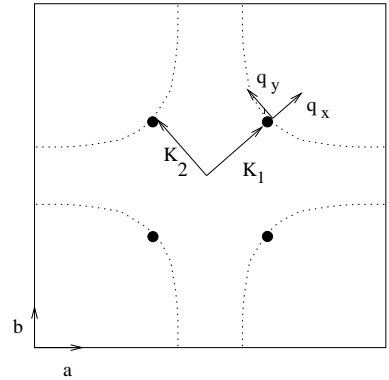}}
\vspace*{1em}

\noindent
\caption{The wavevectors $\vec{K}_i$, $i=1,2$, and $\vec{q}$. The dashed
line stands for the putative Fermi surface. The SDW ordering wave vectors
are $\vec{Q}_{i} = 2 \vec{K}_{i}$. }
\label{f2}
\end{figure}

\section{Coupling to topological defects}

The goal in this section will be 
to find the most economical form of the coupling between
nodal excitations in the dSC and the fluctuations of the phase of the
superconducting order parameter. Working assumption is that
the amplitude fluctuations are frozen well below the
pseudogap  temperature $T^*$, so it is only the phase degree of
freedom that remains active at low energies. With this in mind I write
\begin{equation}
v_\Delta \rightarrow
v_\Delta (\vec{r},\tau) = |v_\Delta| e^{i (\phi_s(\vec{r},\tau)
 +\phi_r (\vec{r},\tau) )},
\end{equation}
where $\phi_r$ represents the regular ("spin-wave") part of the
order parameter phase,
and $\phi_s$ is the singular contribution due to topological defects.
At $T=0$ these would be the vortex loops \cite{nguyen},
or the more familiar vortices and antivortices at $T\neq 0$.
At this point it is
tempting to transform both spin-up and spin-down fermionic operators by
absorbing a half of the total superconducting
phase into each. In presence of topological
defects, however, this would lead to
multivalued fermionic fields and would not be a local change of variables
in the partition function. This problem may be circumvented by 
allowing only vortices of double vorticity \cite{balents}, for example, 
which then leads to the $Z_2$ gauge theory representation
of the problem, and a
possibility of spin-charge separation in the pseudogap regime \cite{senthil}.
It is the single vortices, however, that first become relevant
at the $T\neq 0$ Kosterlitz-Thouless transition \cite{kosterlitz},
and they should be included in the description of the 
$T=0$ transition as well.  I will therefore utilize
 the idea of Franz and Te\v sanovi\' c \cite{marcel}, \cite{anders}
who suggested dividing a given vortex configuration 
into two groups A and B, and transforming the electron
operators with spin up and spin down differently. We write
\begin{equation}
\phi_A (\vec{r},\tau) = \frac{\phi_r (\vec{r},\tau)}{2} +\phi_{sA}
(\vec{r},\tau)  , 
\end{equation}
and similarly for B. $\phi_{sA}$ is the piece of the singular part of the
phase that comes from the defects grouped in  A. 
One may then make a {\it local} change of variables
by introducing a new Dirac field  $\Psi$ as 
\begin{equation}
\Psi (\vec{r}, \tau) = U(\vec{r},\tau) \Psi ' (\vec{r},\tau)
\end{equation}
where $U= diag
\{ e^{-i\phi_A},e^{i\phi_B}, e^{-i\phi_A}, e^{i\phi_B} \}$. 
Since any given vortex defect is either in group A or B,
and therefore associated either with up, or with down spin by the
transformation (9), circling around it with the transformed fermion 
would yield either $2\pi$ or zero of the accumulated phase change.
Components of the new field $\Psi$ are therefore single-valued functions.

   The gauge-transformed action for the Dirac field $\Psi$ is then
\begin{eqnarray}
S[\Psi' ] \rightarrow S[\Psi, \vec{a}, \vec{v}]  = \\ \nonumber
\int d^2 \vec{r} \int_0 ^ {\beta} d\tau \bar{\Psi} _1 [
\gamma_0 (\partial_\tau + i a_0)
+ \gamma_1 v_f (\partial_x + i a_x)  \\ \nonumber 
+ \gamma_2 | v_{\Delta}|  (\partial_y+ i a_y) ] \Psi_1 +
( 1 \rightarrow 2, x\leftrightarrow  y) 
+ i v_\mu J_{\mu}, 
\end{eqnarray}
with $a_\nu =\partial_\nu (\phi_A -\phi_B)/2$,
$v_\nu =\partial_\nu (\phi_A + \phi_B)/2$, and $J_{\nu} =
(\Psi^\dagger_i (I\otimes \sigma_3) \Psi_i ,
v_F \Psi^\dagger_1 (\sigma_3\otimes I ) \Psi_1,
v_F \Psi^\dagger_2 (\sigma_3\otimes I ) \Psi_2)$. Since the
vector $J_\nu$ is built out only of the products of the creation and the
annihilation operators with same spin,
it also represents the {\it physical} charge current carried by the
quasiparticles.  On the other hand, since the regular part of the
phase $\phi_r $ was in the Eq. (8) divided equally between spin up
and spin down,
the Dirac field $\Psi$ is invariant under a regular gauge transformation.
Components of $\Psi$ therefore create electrically
{\it neutral} excitations with spin-1/2 \cite{balents}, 
which may therefore be referred to as {\it spinons}.

The action (10) has two rather different gauge symmetries, and it may be 
worthwhile pausing  a little to reflect on them.
First, the physical electromagnetic gauge field
$A_\mu$ would enter the action (10) by the  replacement
$v_\mu \rightarrow v_\mu + A_\mu$, and couple to the charge current.
Under a regular gauge transformation $A_\mu \rightarrow
A_\mu + \partial_\mu \chi $, the Volovik's field \cite{volovik}
$v_\mu \rightarrow v_\mu
-\partial_\mu \chi $, while the gauge field $a_\mu$ 
 and $\Psi$ remain the same. The
action (10) is therefore gauge invariant, in the standard sense.
But it also has an additional internal gauge symmetry,
under the transformation $a_\mu
\rightarrow a_\mu + \partial_\mu \chi$, $v_\mu \rightarrow v_\mu$,
$\Psi \rightarrow e^{- i \chi} \Psi$. This reflects the freedom of 
choice in the Eq. (8); one could have equally well chosen the
regular part of $\phi_A$ to be $(\phi_r / 2) + \chi$, and of $\phi_B$, 
$ (\phi_r /2) - \chi $. One deals with this gauge freedom,
as usual, by eventually introducing the
gauge-fixing term for $a_\mu$ that allows one to  freely 
sum over all regular internal gauges $\chi$. Similarly, the division
of the singular part of the superconducting
phase into that which comes from  the 
defects in the group A, and the defects in the group B, is equally arbitrary.
Just like one effectively sums over all  {\it regular} internal
gauges by the
introduction of the gauge-fixing term, we will sum over all
{\it singular} internal gauges by averaging over all possible divisions
of defects into two groups. This is explained in the next section,
and in the Appendix B. As a byproduct, the averagings over
regular and singular internal gauges  will 
insure that up and down spinons are treated equally in the
$QED_3$, in respect of the 
symmetry of the original electronic action (2).

The crucial observation about the action (10) is
that the coupling of spinons
to phase fluctuations is furnished by {\it two} U(1) 
fields which play quite different roles in the problem.
Total superconducting phase
determines the Volovik's field
$v_\nu$ and couples to the charge current, in the
same way as the true electromagnetic field would.
$v_\nu$ will therefore inevitably become {\it massive} once the high-energy
spinons in the Eq. (10) are begun to be integrated out. Its
fluctuations therefore may provide only a short-range interaction between
spinons. The gauge field $a_\nu$, on the other hand,
enters (10) in a gauge-invariant way, and therefore is 
protected from acquiring a mass from spinons. Both gauge fields, however,  
depend on the fluctuating positions of the topological defects, and
acquire their dynamics not only from the spinons, but from the defects as well.
To determine their dynamics one therefore
needs to integrate the defect degrees
of freedom out. {\it If} $a_\mu$ would stay 
massless even after this integration is performed,   
it would mediate a 
long-range interaction between the nodal excitations, which,
unlike the short-range quartic terms in the Eq. 6, would not be made
irrelevant by the phase space restrictions. This, however,
depends on the precise way $a_\mu$ acquires its dynamics from  
the fluctuating vortex loops, to which I turn next.

\section{Dynamics of the gauge-fields}

The zero-temperature
partition function for the coupled system of d-wave quasiparticles
and  superconducting phase fluctuations is therefore 
\begin{equation}
Z= \int D[\Psi, \vec{a}, \vec{v}] e^{-( S[\Psi, \vec{a},\vec{v}] + 
S_{U(1)} [\vec{a}, \vec{v}] ) },
\end{equation}
with $S[\Psi, \vec{a}, \vec{v}]$ defined by the Eq. (10), and with 
$S_{U(1)} [\vec{a}, \vec{v}]$ to be derived by integrating
out the phase fluctuations.  For simplicity, I will assume that these
may be described the $2+1$ dimensional XY model. The bare stiffness for the
phase fluctuations will be assumed to be provided by the
high energy modes that have been integrated out in arriving at the
low-energy theory.  Our goal will be then to
rewrite the partition function for the XY model as the functional
integral over the fields $\vec{a}$ and $\vec{v}$. In particular,
we want to integrate over the topological defects implicit in the
XY model. 

I first discretize the space and the imaginary time in writing the
partition function of the XY model. This is done to facilitate
a more rigorous treatment of the topological defects, and it will prove
possible to return to the continuum description we employed
until now. On a lattice, in the standard lattice gauge-theory notation
\cite{nguyen} 
\begin{equation}
Z_{xy} = \int_0^{2\pi} (\prod_i  d\phi_i) \exp (K \sum_{i, \hat{\mu} =
\hat{x},\hat{y},\hat{\tau} }
\cos(\phi_{i+\hat{\mu} }  -\phi_i)),   
\end{equation}
where the index $i$ labels the sites of a three ($2+1$)
dimensional lattice, and
$\hat{x}$ is the lattice unit vector in the $x$ direction, for
example. For simplicity,
full isotropy in the XY model is assumed. Using the Villain
approximation \cite{villain}
and then integrating over the phases leads to 
\begin{eqnarray}
Z= \int_{-\infty}^{\infty} d\vec{s}
\sum_{\vec{n}} '  \exp (-\frac{1}{2K} \sum_i(\nabla \times \vec{s}_i)^2 \\ \nonumber
+ i 2 \pi \sum_i  \vec{n}_{i} \cdot \vec{s}_i ) ,
\end{eqnarray}
where $\vec{n}_i =(n_{i,\tau}, n_{i,x}, n_{i,y})$
is an integer vortex-loop vector variable,
satisfying the constraint $\nabla \cdot \vec{n}_i =0$ (indicated with the prime
on the sum). $\nabla$ and $\nabla \times $ should
be understood as the lattice gradient and the curl,
respectively.  Summing over $\vec{n}_i $  forces $\vec{s}_i $ to take
integer values, and the above expression becomes
the standard current representation of the XY model \cite{nguyen}.

  Next, I imagine dividing
a given configuration of vortex loops into two arbitrary groups,
and write $\vec{n}_i =\vec{n}_{A,i} + \vec{n}_{B,i}$,
with $\nabla \cdot \vec{n}_{A,i} = \nabla \cdot \vec{n}_{B,i} =0$.
We will want to sum over
all integer $\vec{n}_{A,i}$ and $\vec{n}_{B,i}$,
in order to average over all
possible divisions of vortices into two groups.
Introducing the lattice version of
the fields $\vec{a}_i$ and $\vec{v}_i$ as $\vec{B}_i+\vec{b}_i
 = 2\pi \vec{n}_{A,i}$,
$\vec{B}_i -\vec{b}_i = 2\pi \vec{n}_{B,i} $, where 
$\vec{b}_i= \nabla \times \vec{a}_i$ and $\vec{B}_i= \nabla \times \vec{v}_i$,
I write \cite{rem}
\begin{eqnarray}
Z_{xy} = \int_{-\infty}^{\infty} d[ \vec{a},\vec{v}, \vec{t},\vec{s}, \vec{r}]
\sum_{\vec{n}_A, \vec{n}_B} ' 
\exp -\sum_i [ \frac{1}{2K} (\nabla \times \vec{s}_i) ^2 \\ \nonumber
+ i 2\pi \vec{s}_i \cdot (\vec{n}_{A,i} + \vec{n}_{B,i} ) \\ \nonumber 
+ i \vec{t}_i \cdot (\vec{B}_i +\vec{b}_i - 2 \pi \vec{n}_{A,i} ) 
+ i \vec{r}_i (\vec{B}_i-\vec{b}_i - 2 \pi \vec{n}_{B_i} )] . 
\end{eqnarray}
The summations over 
$\vec{n}_{A,i}$ and $\vec{n}_{B,i} $ then enforce the constraints 
$\vec{s}_i -\vec{t}_i = \vec{m}_{A,i} $, and $\vec{s}_i - \vec{r}_i
= \vec{m}_{B,i} $, where $\vec{m}_{A,i} $ and $\vec{m}_{B,i}$ are
new integers. Performing the Gaussian integrals over $\vec{s}_i$, yields
\begin{eqnarray}
Z_{xy} = \int_{-\infty}^{\infty} d [\vec{a},\vec{v}]
\sum_{\vec{m}_A ,\vec{m}_B } '
\exp -\sum_i  [ 2 K (\nabla \times \vec{v}_i )^2 +\\ \nonumber
 i \vec{v}_i \cdot (\nabla \times (\vec{m}_{A,i} + \vec{m}_{B,i} )) 
+ i\vec{a}_i \cdot (\nabla\times ( \vec{m}_{A,i} - \vec{m}_{B,i} ) ) ].
\end{eqnarray}
This can be further simplified by noticing that the action is quadratic in the
Volovik's field $\vec{v}$, which can also be integrated out. In doing so I will
neglect the additional coupling of $\vec{v}$ to the charge current $\vec{J}$ in the
Eq. (10), which only leads to additional irrelevant interaction between spinons. The
integration over $\vec{v}_i$ in the last equation then gives
\begin{eqnarray}
Z_{xy} = \int_{-\infty}^{\infty}
d\vec{a}  \sum_{\vec{m}_A ,\vec{m}_B } ' 
\exp -\sum_i  [ \frac{1}{8K} \\ \nonumber 
(\nabla \times (\vec{m}_{A,i}+\vec{m}_{B,i})) ^2
+ i\vec{a}_i \cdot (\nabla\times ( \vec{m}_{A,i} - \vec{m}_{B,i} ) ) ]
\end{eqnarray}
Integrating over $\vec{a}_i $ in (16) would give back the current
representation of the XY model, Eq. (13). 
Alternatively, we can introduce the real variables $\vec{\Phi}_{+,i}$ and
$\vec{\Phi}_{-,i}$ and write 
\begin{eqnarray}
Z_{xy}= \int_{-\infty}^{\infty} d[ \vec{a},\vec{\Phi}_-,\vec{\Phi}_+]
\sum_{\vec{l}_A, \vec{l}_B} ' 
\exp -\sum_i ( \frac{1}{8K} (\nabla \times \vec{\Phi}_{+,i} )^2\\ \nonumber
+ i \vec{a}_i \cdot (\nabla \times \vec{\Phi}_{-,i} )  +
i 2\pi ( \vec{l}_{A,i}\cdot \vec{\Phi}_{A,i}
+ \vec{l}_{B,i} \cdot \vec{\Phi}_{B,i}) )
\end{eqnarray}
where $\vec{\Phi}_{+,-,i} = \vec{\Phi}_{A,i} \pm  \vec{\Phi}_{B,i}$.
The summations over the auxiliary link variables
$\vec{l}_{A,B}$ force $\vec{\Phi}_{A}$ and $\vec{\Phi}_B$,
and therefore $\vec{\Phi}_{+}$ and $\vec{\Phi}_{-}$
to be integers. To preserve the gauge invariance
($\Phi_{+, i, \mu} \rightarrow \Phi_{+,i,\mu} + \nabla_{\mu}\chi_{+,i} $,
$\Phi_{-, i, \mu} \rightarrow \Phi_{-,i,\mu} + \nabla_{\mu} \chi_{-,i}$) 
of the last expression we must impose
$\nabla\cdot \vec{l}_{A,i}=\nabla\cdot\vec{l}_{B,i}= 0$
\cite{fisherlee}, \cite{herbut4}.
We may next add a small chemical potential for the
link variables $\vec{l}_{A,B}$ to the action in Eq. (17) as the term 
$x \sum_i (\vec{l}_{A,i} ^2 + \vec{l}_{B,i} ^2 )$.  Up to the 
Villain approximation, the last expression is then equal to
\begin{eqnarray}
Z_{xy} = \lim_{x \rightarrow 0} \int_{-\infty}^{\infty}
d[ \vec{a},\vec{\Phi}_{A}, \vec{\Phi}_{B}] \int_{0}^{2\pi} d[
\theta_{A}, \theta_{B}] \\ \nonumber
\exp -\sum_i [ \frac{1}{8K} (\nabla \times \vec{\Phi}_{+,i} )^2
+ i \vec{a}_i \cdot (\nabla \times \vec{\Phi}_{-,i})  \\ \nonumber
-\frac{1}{2x} \cos( \theta_{A, i}-\theta_{A,i+\hat{\nu}}
-2 \pi \Phi_{A, i, \hat{\nu}})   \\ \nonumber 
-\frac{1}{2x} \cos( \theta_{B, i}-\theta_{B,i+\hat{\nu}}
-2 \pi \Phi_{B, i, \hat{\nu}}) ], 
\end{eqnarray}
where I introduced two sets of "dual" angles $\theta_{A,i}$ and
$\theta_{B,i}$ to insure the 
gauge invariance, and imposed the "frozen" limit $x\rightarrow 0$.  The 
integration over $\vec{a}_i$ in the Eq. (18) together with the frozen limit
ultimately sets $\theta_{A,i} \equiv \theta_{B,i}$,
so the last equation becomes
another representation of the frozen lattice superconductor (FLS),
which is well known to be dual to the XY model in three dimensions
\cite{peskin}, \cite{herbut3}.

In principle, one would like to integrate out all the fields  other than
$\vec{a}$ in the Eq. (18), to be left with the effective action
$S_{U(1)}[\vec{a}]$ for $\vec{a}$ only. The result
would be an interacting theory for $\vec{a}$,
which can be expanded in powers of $\vec{a}$, for example.
Instead of doing this, I will approximate the 
$S_{U(1)}[\vec{a}]$ with the effective Gaussian action for $\vec{a}$, that 
reproduces the gauge-field propagator in the full theory (18).
This approximation
may  be understood as the self-consistent mean field theory for $\vec{a}$,
with the effect of integration over
all other fields in (18) lumped into the form of the
propagator.

In this approximation the problem of dynamics of the gauge-field $\vec{a}$
reduces to the computation of the two-point correlation function for
$\vec{a}$ from the representation of the XY model in Eq. (18). I therefore  
introduce the source term into the last expression by
 adding $i \sum_i  \vec{j}_i \cdot (\nabla\times \vec{a}_i ) $
 to the exponent. Then
 \begin{equation}
\langle (\nabla\times\vec{a})_{i, \nu}(\nabla\times\vec{a})_{j, \mu}\rangle
= \frac{\partial^2}{\partial j_{i,\nu} \partial j_{j,\mu}}
\ln Z_{xy}|_{\vec{j}\equiv 0}. 
\end{equation}
It is convenient then to  integrate over $\vec{a}$ in the $Z_{xy}$
first. One finds
 \begin{eqnarray}
\langle (\nabla\times\vec{a})_{i,\nu}(\nabla\times\vec{a})_{j,\mu}\rangle
=  \delta_{i,j} \delta_{\nu,\mu}  \lim_{x \rightarrow 0} \\ \nonumber 
\frac{\pi ^2 }{x}
\langle \cos(\theta_i - \theta_{i+\hat{\nu}}
-2 \pi \Phi_{i,\nu}) \rangle_{FLS},
\end{eqnarray}
where the last average is to be taken over the configurations
of the FLS 
\begin{eqnarray}
Z_{xy} = \lim_{x\rightarrow 0} \int d[\vec{\Phi},\theta] 
\exp - \sum_{i, \hat{\nu}}
 [ \frac{1}{2K} (\nabla \times \vec{\Phi})^2 \\ \nonumber
- \frac{1}{x} ( \cos( \theta_{i}-\theta_{i+\hat{\nu}}
-2 \pi \Phi_{i, \hat{\nu}}) ] . 
\end{eqnarray}

It is well established that the lattice superconductor at a small but finite
"temperature" $x$ has a phase transition as $K$ is varied
in the same universality class as in the frozen limit $x=0$
\cite{nguyen}, \cite{peskin}, \cite{olson}.
We may therefore relax the constraint $x\rightarrow 0$ with impunity
and assume $x$ to be finite. The average that appears
on the right hand side of the Eq. (20) can then be computed, for example,
by using the mean-field approximation to the FLS action (21) 
(see Appendix A). This yields
\begin{equation}
\frac{1}{x} \langle \cos(\theta_i - \theta_{i+\hat{\nu}}
-2\pi \Phi_{i,\nu}) \rangle_{FLS}\propto
|\langle\exp (i \theta_i) \rangle |^2.
\end{equation}
This result is quite general, and it simply expresses the fact
that in the ordered phase of the theory  (21) the dual angles
become correlated, while at the same time 
the gauge-field becomes massive via Meissner effect. The gauge field
fluctuations can then be neglected, which makes the requisite
average finite when the dual angles $\theta$ order, i. e. in the
{\it disordered} phase of the original XY model.

Returning to the continuum notation, and switching to  the Fourier space,
the gauge-invariant expression for the
correlation function (19) at low momenta is therefore
\begin{equation}
\langle (\nabla\times\vec{a})_{\nu}(\nabla\times\vec{a})_{\mu}\rangle
\propto (|\langle \Phi \rangle |^2 +O(q^2))  (\delta_{\mu \nu} - \hat{q}_\mu
\hat{q}_{\nu}), 
\end{equation}
where I allowed, in general, for some momentum dependence (the term $O(q^2)$).
$O(q^2)$ term should be expected to appear in a
more sophisticated approximation for the gauge-field dynamics than
provided by the Eq. (20). To the lowest order, the integration over all other
fields in (18) effectively yields the {\it Maxwell term} for the 
gauge field $\vec{a}$, with the stiffness inversely
proportional to the expectation value of the
{\it dual} loop condensate $\langle \Phi \rangle \sim \langle e^ {i\theta }
\rangle $ that reflects the phase of the XY model.
This is the main result of this section.
When the dSC is phase coherent and the
vortex loops are finite in size, $\langle \Phi \rangle =0$, 
and $\vec{a}$ is infinitely stiff,
and in first approximation may be considered 
decoupled from spinons. When vortex
loops blow up, $\langle \Phi \rangle \neq 0$, phase coherence is lost, 
and the spinons are minimally coupled to a massless gauge field.
This is in agreement with the physical arguments advanced in \cite{franz}.

 At high temperature one can neglect the fluctuations in
the imaginary time direction
and deal with the purely 2D problem of point vortices and
antivortices. This simplifies the analysis in that
no gauge invariance needs to be insured in the
Eq. (17), so no dual angles are
required \cite{fisherlee},
\cite{herbut4}. One then ends up with the thermodynamic
vortex  fugacity playing the role of the dual condensate \cite{herbut2},
and with the simpler sine-Gordon theory instead of the FLS.
For an alternative derivation of the gauge field dynamics  
at $T\neq 0$ and in continuum
that is in full accord with the conclusions of this
section I direct the reader to the Appendix B.

There is an additional subtlety in going from the lattice to
the  continuum theory that is worth registering.
The partition function in the Villain approximation for the
XY model in the Eq. (17) has the symmetry under $a_{i,\mu} \rightarrow
a_{i,\mu} + 2 \pi n_{i,\mu}$, with $n_{i,\mu}$ integer,
that becomes broken when a small chemical potential $x \neq 0$ for
the link variables $\vec{l}_{A,B}$ (in passing to the Eq. (18)) is added.
This periodicity would dictate
 that the summation over the integer vortex variables
in Eq. (17) with $x=0$ should yield a {\it compact}
term for $\vec{a}$. Absence of the chemical potential $x$, i. e. of the
vortex core energy, in the Eq. (17), on the other hand, must be regarded as
an artifact of the Villain representation to the original XY model,
in which, as well known, vortices do cost finite energy \cite{kosterlitz}.
This is because the Villain approximation reproduces correctly only the
long-range part of the vortex interaction, while the short range part
needs to be modified "by hand" \cite{jose}
in order to obtain the finite core energy.
The dynamics of $\vec{a}$ should therefore be determined from the
theory with $x\neq 0$, and by the
{\it non-compact} Maxwell term, as in the Eq. (23). 
Possible effects of compactness of
$\vec{a}$ on the picture developed in this paper are discussed
in the sec. XII.

\section{Dynamical breaking of chiral symmetry}

  The effective $T=0$ low-energy 
theory for the interacting system of d-wave quasiparticles and fluctuating
vortex loops, after the integration over vortex loops is therefore
 \begin{eqnarray}
S[\Psi] = \int d^2 \vec{r} d\tau \{  \bar{\Psi} _1 [
\gamma_0 (\partial_\tau + i a_0)
+ \gamma_1 v_f (\partial_x + i a_x)  \\ \nonumber 
+ \gamma_2 | v_{\Delta}|  (\partial_y+ i a_y) ] \Psi_1 +
( 1 \rightarrow 2, x\leftrightarrow y ) \\ \nonumber
+ \frac{1}{2|\langle \Phi \rangle |^2 }
( c^2 (\nabla \times \vec{a})^2 _ \tau +
(\nabla \times \vec{a})^2 _ {\vec{r}} ) \},  
\end{eqnarray}
where I omitted the higher derivative terms, and the terms
 quartic in $\Psi$. This is the standard three dimensional quantum
electrodynamics ($QED_3$), with two important caveats: 1) the
coordinates $x$ and $y$ are exchanged for the second Dirac field,
2) there is an inherent anisotropy in the model, $v_f\neq
v_\Delta\neq c $, where $c$ is a characteristic velocity for
the phase fluctuations \cite{randeria}.
First, let us consider the simpler isotropic
limit of the theory, $v_f = v_\Delta=c $. There are sixteen $8\times 8$
matrices then that either commute or anticommute with the three $8\times 8$
$\gamma$-matrices that appear in the Eq. (24):
$diag \{ \gamma_0, \gamma_0 \}$, $ diag\{\gamma_1, \gamma_2 \}$,
$diag \{\gamma_2, \gamma_1 \}$.
 First, there are eight block-diagonal Hermitean matrices
\begin {equation}
I\otimes I_4, \sigma_3 \otimes I_4, I\otimes \gamma_{35}, \sigma_3 \otimes
\gamma_{35}, 
\end{equation}
that commute, and
\begin{equation}
I\otimes \gamma_3, \sigma_3 \otimes \gamma_3, I\otimes \gamma_5,
\sigma_3 \otimes \gamma_5, 
\end{equation}
that anticommute with the $\gamma$-matrices.
Here, $\gamma_3 = \sigma_2 \otimes \sigma_2 $, $\gamma_5 = \sigma_3 \otimes
I$, $ \gamma _{35} = i \gamma _3 \gamma_5$, and
$I_4 = I \otimes I$. Next, there are eight more
block-off-diagonal Hermitean matrices
\begin{eqnarray}
\sigma_1 \otimes \frac{i}{\sqrt{2}} (\gamma_2 - \gamma_1)\gamma_3,
\sigma_2 \otimes \frac{i}{\sqrt{2}} (\gamma_2 - \gamma_1)\gamma_3, \\ \nonumber
\sigma_1 \otimes \frac{i}{\sqrt{2}} (\gamma_2 - \gamma_1)\gamma_5,
\sigma_2 \otimes \frac{i}{\sqrt{2}} (\gamma_2 - \gamma_1)\gamma_5,
\end{eqnarray}
that commute, and
\begin{eqnarray}
\sigma_1 \otimes \frac{1}{\sqrt{2}} (\gamma_1 - \gamma_2),
\sigma_2 \otimes \frac{1}{\sqrt{2}} (\gamma_1 - \gamma_2), \\ \nonumber
\sigma_1 \otimes \frac{i}{\sqrt{2}} \gamma_0 (\gamma_1 + \gamma_2),
\sigma_2 \otimes \frac{i}{\sqrt{2}} \gamma_0 (\gamma_1 + \gamma_2),
 \end{eqnarray}
that anticommute with the $\gamma$-matrices.
I call these sixteen generators $G_i$, $i=1, ... 16$, in the above order.
The isotropic $QED_3$ in the Eq. (24) is invariant under a global 
unitary transformation
\begin{equation}
\Psi \rightarrow U \Psi , 
\end{equation}
where
\begin{equation}
U= e^{ i \sum_{i=1}^{16} \theta_i G_i } . 
\end{equation}
This follows immediately by observing that all the generators
commute with the $8\times 8$
matrices $ diag \{ \gamma_0,\gamma_0 \} diag \{ \gamma_1,\gamma_2 \}$,
and  $ diag \{ \gamma_0,\gamma_0 \} diag \{ \gamma_2,\gamma_1 \}$
by construction. The
unitary transformations in (29) can be shown to
form the Lie group $U(4)$. Following the standard terminology in the
field theory literature, I will refer to this symmetry of the $QED_3$ as 
"chiral".

As a first step towards understanding of the meaning of
the  chiral symmetry in the present context, it will prove 
useful to consider how it may be broken. 
$QED_3$ is well known to have the  chiral
symmetry spontaneously broken \cite{appelquist}, by dynamical  
generation of the mass term in the action (24):
\begin{equation}
m \int d^2 \vec{r} d\tau
\sum _{i=1}^{2} \bar{\Psi}_i \Psi_i,
\end{equation}
with $m\propto |\langle \Phi \rangle|^2 $, i. e. proportional to the
effective charge of the $QED_3$. Containing just a single
$\gamma$-matrix, the mass term in the Eq. (31)
breaks all the {\it anticommuting}
generators, $G_i$ with $i= 5,6,7,8,13,14,15,16$.
The chiral symmetry is reduced from $U(4)$ to $U(2)\times U(2)$, with eight
generators preserved. The fermion mass is generated dynamically
due to the coupling to the gauge field. To see this,
neglect the wave-function renormalization and
the vertex corrections (which can be rationalized in the limit of a large
number of Dirac fields N),  and write the self-energy as
\begin{equation}
\Sigma(q) = |\langle \Phi \rangle |^2 \gamma_\nu
\int \frac{d^3 \vec{p}} {(2\pi)^3 } \frac{D_{\nu \mu}  (\vec{p} -\vec{q})
\Sigma(p)}{ p^2 + \Sigma^2 (p)} \gamma_\mu,
\end{equation}
where $\vec{q} = (\omega,q_x,q_y)$. The gauge-field propagator in the
transverse (Landau) gauge is
\begin{equation}
D_{\nu \mu}  (\vec{p} )= (\delta_{\nu\mu} - \hat{p}_\nu \hat{p}_\mu ) /
(p^2 + \Pi (p)), 
\end{equation}
where $\Pi(p)$ is the self-consistently computed polarization. At
$p\ll \Sigma(0)=m $, assuming  a finite mass $m$ gives
\begin{equation}
\Pi(p) = \frac{N |\langle \Phi\rangle|^2} { 6 \pi} \frac{p^2}{m}   +O(p^4).
\end{equation}
For the polarization at all momenta see the Appendix C.
The Eq. (32) was analyzed in \cite{appelquist} (see also Appendices C and E), 
and there is a solution with finite $m$ for the number of
Dirac fields $N<N_c = 32/\pi^2 = 3.24$.
Full numerical solution that includes the wave-function renormalization
and vertex corrections confirms that $N_c \approx 3$ \cite{maris},
almost independently of the choice of vertex. Lattice simulations give
$3< N_c <4$ \cite{kocic}, or at least that $N_c >2$ \cite{alexandre}.
It therefore seems reasonable to conclude that 
for $N=2$ the chiral symmetry in the isotropic
$QED_3$ becomes spontaneously broken when the vortex loops unbind
and $\langle \Phi \rangle \neq 0$.

 Since the matrix $\gamma_0$ commutes with the
electron-spinon transformation in the Eq. (9), it is easy to rewrite the 
mass term in the $QED_3$ in terms of the original electron operators: 
\begin{eqnarray}
m \sum _{i=1}^{2} \bar{\Psi}_i \Psi_i \rightarrow  
m T \sum _{\vec{k}\approx \vec{K}_{1},\omega_n}
\{ [c_+ ^\dagger (\vec{k}, \omega_n) c_+ (\vec{k}-\vec{Q}_1,\omega_n) \\ \nonumber
- c_- ^\dagger (-\vec{k}+\vec{Q}_1, \omega_n) c_- (-\vec{k},\omega_n) ] +
\\ \nonumber 
[c_+ ^\dagger (\vec{k}-\vec{Q}_1, \omega_n) c_+ (\vec{k},\omega_n) \\ \nonumber 
- c_- ^\dagger (-\vec{k}, \omega_n) c_- (-\vec{k}+\vec{Q}_1,\omega_n) ] \}
+ (1\rightarrow 2).  
\end{eqnarray}
The reader will recognize this as the low-energy part of the
{\it staggered} potential along the spin $z$-axis
\begin{equation}
m \int d^2 \vec{r}d\tau \sum_{\sigma=\pm, i=1,2}
\cos( \vec{Q}_i \cdot \vec{r}) \sigma c_\sigma ^\dagger (\vec{r},\tau)
c_\sigma (\vec{r},\tau), 
\end{equation}
so the mass in the $QED_3$ is nothing but the spontaneously generated  
SDW order parameter. The periodicity of the SDW 
is set by the vectors $\vec{Q}_i$, and thus tied to the Fermi surface.
The SDW order is established as soon as the phase coherence is lost,
and the charge $ \langle \Phi\rangle \neq 0$.
In the large-N approximation \cite{appelquist} one finds that
\begin{equation}
m \approx 16 
 |\langle \Phi \rangle|^2 e^{ -2 \pi /\sqrt{ ( \frac{N_c}{N} -1) } }.  
\end{equation}
Since $ N_c \approx 3$, for $N=2$ one finds
that $m\sim 10^{-2} |\langle \Phi \rangle|^2 $. This
extreme "lightness" of  fermions in the $QED_3$
derives from the fact that the mass comes solely from
the interaction with the soft gauge-field.

Breaking of chiral symmetry in the $QED_3$ also implies that the
energies of spinons have become complex and finite
in the phase incoherent state with $\langle \Phi \rangle \neq 0$.
In the simplest approximation the electron
propagator may be computed as a product of the spinon
and the gauge-field
propagators, so a spinon "gap" should imply a charge gap as well, i. e. the
system becomes an insulator \cite{khvesch}.
In section IX I discuss how
spinons should actually be confined in the insulating state.
Staggered magnetization, charge gap, and the spinon confinement
when taken together imply that the state with broken chiral
symmetry is nothing but the standard, albeit a weak, SDW.
It seems natural to assume then that this state is continuously
connected to the antiferromagnet near half-filling in cuprates.
This expectation is further corroborated by considering the effect 
of Coulomb interactions, which is done in section VIII. 

 It has been already mentioned that we have some freedom in choosing
 the representation  of the $\gamma$-matrices. In particular, it was the
 specific choice of $\gamma_0$ that led to the cos-SDW order
 parameter displayed in the Eq. (35-36).
 In the next section I discuss how "rotating" the cos-SDW by the broken
 chiral generators leads to a different insulating states.

 \section{More on chiral symmetry: the space of insulators }

 In discussing the pattern of chiral symmetry breaking 
 in the $QED_3$ one needs to distinguish at least two different cases. The 
 isotropic theory  ($v_\Delta = v_f$) has the full $U(4)$ symmetry in
 its massless phase, so the mass term breaks eight of its sixteen
 generators. In cuprates \cite{chao},
  however,  $v_f/v_\Delta \sim 10 $, 
 and even with $m=0$ the symmetry is only $U(2)\times U(2)$, generated
 by the block-diagonal $G_i$ $i=1,...8$. 
 How such a large anisotropy affects the value of $N_c$ is a non-trivial
  problem, and is addressed in a separate publication \cite{dom}. 
 Here I will consider only the effect of anisotropy on the chiral 
 symmetry, and assume it is reduced to $U(2) \times U(2)$. It suffices 
 then to look at each Dirac component in the $QED_3$ separately, i. e.
 consider just the $4 \times 4$ representation of the $\gamma$-matrices,
 as defined right below the Eq. (5).

 It can be easily shown that any matrix
that anticommutes with both $M_1$ and $M_2$ and squares to unit matrix 
may be chosen as $\gamma_0$, and will  lead to a representation of 
the $\gamma$-matrices like in the Eq. (6). The mass term
$\sim m\Psi^\dagger \gamma_0 \Psi$ in the action would then gap the 
quasiparticles, in analogy to the standard relativistic Dirac equation.
The problem of different chiral orders is therefore nothing else
but finding all the ways in which d-wave quasiparticles can spontaneously
acquire such a "relativistic mass". It will be useful to introduce  
"directions" in the space of broken symmetry states, as
a set of linearly independent matrices that anticommute with
$M_1$ and $M_2$, and square to one. It is easy to show that in  the 
$4\times 4$ representation there are only four such matrices
\begin{equation}
\tilde{\gamma}_0, \tilde{\gamma}_3, \tilde{\gamma}_5,
i\tilde{\gamma}_1 \tilde{\gamma}_2,
\end{equation}
with  $\tilde{\gamma}_0 = \gamma_0$, and where $\tilde{\gamma}_1 = -i M_1$,
$\tilde{\gamma}_2 = iM_2$, $\tilde{\gamma}_3 = \sigma_3 \otimes \sigma_2$,
and $\tilde{\gamma}_5 = \sigma_2 \otimes I$.  In principle, any of these
four if used instead of our $\gamma_0$ in the construction
of the Dirac theory in the Eq. (6) and in the mass term would 
give a relativistic gap to Dirac fermions.  The last matrix,
\begin{equation}
i\tilde{\gamma}_1 \tilde{\gamma}_2 = I\otimes \sigma_2, 
\end{equation}
however, being a product of two $\tilde{\gamma}$-matrices
does not break the chiral symmetry,
and is believed not to be spontaneously generated in the
$QED_3$ \cite{appelquist}, \cite{vafa}.
I therefore focus on the remaining three.
Choosing one among $\{ \tilde{\gamma}_0,\tilde{\gamma}_3,\tilde{\gamma}_5
\} $ as the $\gamma_0$-matrix in the mass term
reduces the $SU_c (2)$ subgroup of $U(2)(=U(1)\times SU_c (2)) $,
generated by $\{\gamma_3,\gamma_5,\gamma_{35} \}$, 
to $U_c (1)$. The two anticommuting generators of the
$SU_c (2)$ that are broken then 
rotate the chosen order parameter towards the two remaining
"directions" in the chiral space. For example: for our choice of
$\tilde{\gamma}_0 = \gamma_0$, it is $\gamma_{35}$ that remains
unbroken in the cos-SDW phase, whereas the broken generators
rotate the cos-SDW order parameter as 
\begin{equation}
e^{i \theta \gamma_i} \tilde{\gamma}_0 e^{-i \theta \gamma_i} =
\cos(2\theta) \tilde{\gamma}_0 - \sin(2\theta) \tilde{\gamma}_i; i=3,5. 
\end{equation}
Choosing $i=5$, for example, for both Dirac fields 
rotates the cos-SDW in the Eq. (35) into
\begin{equation}
m \int d^2 \vec{r}d\tau \sum_{\sigma=\pm, i=1,2}
\cos( \vec{Q}_i \cdot \vec{r}+ 2\theta)
\sigma c_\sigma ^\dagger (\vec{r},\tau) c_\sigma (\vec{r},\tau).   
\end{equation}
Chiral rotations generated by
$\gamma_5$ thus correspond to {\it sliding} modes of the
SDW. $\gamma_3$, on the other hand, rotates $\tilde{\gamma}_0$ towards the
direction of $\tilde{\gamma}_3$, which describes an additional
particle-particle pairing between the neutral spinons,
with the opposite sign for the diagonally opposed nodes. This may
be understood as an additional p-wave pairing between the spinons, so the
state described by $\tilde{\gamma}_3$ order parameter may be called the
"d+ip" state \cite{zlatko}. 
This state preserves the superconducting $U(1)$ symmetry and the
translational invariance, but breaks the spin-rotational invariance
and is odd under parity. Since $\tilde{\gamma}_3$ {\it does not} commute with
the electron-spinon transformation (9), however, "d+ip" state can not
be that simply expressed in terms of electronic operators, as it
proved possible for the SDW states. 
The relationship between the directions in the order parameter
space  $\{ \tilde{\gamma}_0,\tilde{\gamma}_3,\tilde{\gamma}_5 \} $,
and the chiral generators may be summarized  pictorially
as on Fig. 3. 

\begin{figure}
\centerline{\epsfxsize=7cm \epsfbox{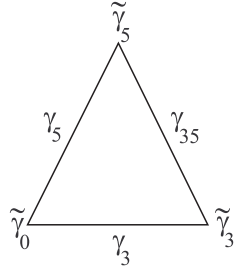}}
\vspace*{1em}

\noindent
\caption{The corners of the triangle represent the
three chiral directions in the space of insulating states that descend
from the d-wave superconductor. At the side opposite to
a particular direction lies the corresponding unbroken chiral
generator, while the remaining two broken generators
rotate the chosen insulator towards two other directions.  }
                                                           
\label{f2}
\end{figure}

It is instructive to look more closely at the origin of the
$U(2)$ symmetry (per Dirac component) 
that appears in the low energy theory of the
dSC. First, the transformations in the
$U(1)$ subgroup of $U(2)=U(1)\times SU_c (2)$ are analogous to the 
spin rotations around the z-axis. To see this, consider the conserved
current that corresponds to the $U(1)$ subgroup: $
J_{i,\mu}= \bar{\Psi}_i \gamma_\mu
\Psi_i $, so that the conserved charge is simply the z-component of the
total spin of the low-energy quasiparticles, one charge
per each pair of nodes. Of course, the quasiparticle action (2) has the full
$SO(3)$ spin symmetry, and this is not to imply that a part of it
is broken in the dSC. It only means that in writing the full action 
(2) in terms of the Dirac fields (3) only the subgroup
of spin rotations around z-axis is represented by a simple
$4\times 4 $ matrices that act on $\Psi$. The rest is still present,
but not that obvious in our choice of the Dirac fields, which was made to
make the chiral symmetry manifest. (For a complementary representation
that is fully rotationally symmetric at the expense of chiral symmetry
see the Appendix D.) The $U(1)$ sub-symmetry is therefore {\it always}
present, both in the superconducting and the insulating states.
The $SU_c (2)$ factor is  more interesting. The conserved currents (per
pair of nodes) in the dSC that correspond to this symmetry are
$J_{i,\mu} ^\Gamma = \bar{\Psi}_i \gamma_\mu \Gamma \Psi_i $, where
$\Gamma = \gamma_3, \gamma_5, \gamma_{35}$. As we have
seen already, the $\gamma_5$ generator simply translates in the
diagonal direction. The corresponding conserved charge may be written as 
\begin{equation}
Q_{i}^5 = \int d \vec{r} d\tau J_{i,0}^5 = T\sum_{\sigma,
\omega_n, \vec{k} \approx \pm \vec{K}_{i} }
\pm  c^\dagger _\sigma(\vec{k},\omega_n) c_\sigma(\vec{k},\omega_n), 
\end{equation}
and may be identified with the quasiparticle momentum along $\vec{K}_i$. 
More precisely, under the translation of the
original electron operators 
$ c_\sigma (\vec{k}, \omega) \rightarrow e^{i\vec{k}\cdot \vec{R}}
c_\sigma (\vec{k}, \omega)$, the spinon field  transforms as
\begin{equation}
\Psi_i (\vec{r},\tau) \rightarrow e^{ i ( \vec{K}_i \cdot
\vec{R} )  \gamma_5 } \Psi_i (\vec{r}+\vec{R},\tau),  
\end{equation}
where $\vec{k}=\vec{K}_i +\vec{q}$. 
The low-energy theory therefore has more symmetry than the original
action (2), as the chiral rotation by $\gamma_5$ and the translations
of $\Psi$ separately are still the symmetries of the theory (24),
while only when combined as above do they represent an
ordinary translation.  Nevertheless, breaking of chiral
generator $\gamma_5$ implies breaking of the translational
symmetry in the theory.  The remaining two
generators of the chiral $SU_c (2)$, $\gamma_3$ and $\gamma_{35}$,
on the other hand, are not related to any spatial symmetry. They 
should be understood as "internal", and
approximate, symmetries of the dSC that emerge at low energies.
They rotate the translationally invariant "d+ip" state into
a SDW, and therefore connect the two fundamentally different types
of insulators.

 The reader should also note that
 in the action (24) the order parameter can be rotated
independently for the first and the second Dirac field. Any
linear combination of $\tilde{\gamma}_0$ $\tilde{\gamma}_3$, and
$\tilde{\gamma}_5$ is a regular
order parameter too. Since $\tilde{\gamma}_5$ is just a sin-SDW, the 
fundamentally different states are just the SDW are $d+ip$
state. This, however, already leads to a variety of insulating 
phases.  For example, one can choose the 
cos-SDW for the first Dirac field ($\vec{Q}_1$), while being in
the $d+ip$ state for the second. This would correspond to a
one-dimensional SDW along one of the diagonals.

    With the velocity
 anisotropy neglected the $QED_3$ has a larger $U(4)$ symmetry,
with sixteen generators $G_i$. The mass term now breaks all eight 
anticommuting generators, and the chiral manifold of insulating
states becomes larger. For instance, rotating the
cos-SDW with $\theta=\pi/4$ and the generator $G= G_{13}-G_{15}$
leads to a uniform state with an additional "s" component of pairing
between spinons, "d+is" \cite{babak}.  Interestingly, rotating the cos-SDW
by block-off-diagonal generators may also lead to {\it charge stripes}. 
For example,
taking $\theta=\pi/4$ and $G_{15}$, rotates the $8\times 8$
cos-SDW order parameter $I \otimes \tilde{\gamma} _0$ into $(1/2) \sigma_1
\otimes (\gamma_1 + \gamma_2) $. When written in terms of the electronic
operators, this order parameter corresponds to the {\it one-dimensional}
charge density-wave with
the periodicity $\vec{P}_b = \vec{K}_2 + \vec{K}_1$, and with
residual pairing correlations in the orthogonal a-axis direction.
It has been known that stripes indeed occur in some 
high-$T_c$ materials \cite{orenstein}. Here they emerge as insulating
cousins of the d-wave state in the isotropic limit of the theory. It is
also interesting that stripes seem always to be accompanied by
the residual pairing correlations, so one can think of them as
weakly coupled one-dimensional systems on the verge of becoming
phase coherent.

\section{Reduction of chiral symmetry by the irrelevant terms}

  We saw that  the low-energy theory of dSC has the chiral $U(2)$
symmetry per Dirac component, which when spontaneously broken leads
to emergence of the SDW or the $d+ip$ insulators. This enlarged 
symmetry arises only at low-energies, and the irrelevant terms
omitted in the Eq. 6 reduce the $U(2)$ to $U(1) \times U_c (1)$. 
In this section I show the higher order derivatives and the Hubbard
repulsion reduce the chiral $SU_c (2)$ symmetry
to just translations, generated by $\gamma_5$. 
However, we will also find that if both
perturbations are weak it will actually be the SDW solution that
is energetically preferred.

  Let us first consider the higher derivative terms in the Eq. (6).
Since $\xi(\vec{k}-\vec{Q}_1) = \xi(\vec{q}-\vec{K}_1)$, and
$\xi(\vec{q}-\vec{K}_1)= \xi(\vec{K}_1-\vec{q})$,
and analogously for $\Delta(\vec{k})$, one can write the
second-order derivatives in the Eq. (6) as 
\begin{eqnarray}
S_1 = -i \int d^2 \vec{r} d\tau
\bar{\Psi}_1 ' \gamma_5 (\gamma_1 \xi '' (\partial^2) - \gamma_2
\Delta '' (\partial^2 ) ) \Psi_1 ' \\ \nonumber
+ (1\rightarrow 2, x\leftrightarrow y),
\end{eqnarray}
where $\xi ''$ and $\Delta ''$ are the functions coming
from the expansion of $\xi(\vec{k})$ and $\Delta(\vec{k})$ around
$\vec{K}_{1(2)}$, respectively. Their specific forms are model
dependent, and will not be of importance here. What is
important is that $S_1$ manifestly breaks the part of the chiral
symmetry generated by $\gamma_3$ and $\gamma_{35}$,
while preserving only the translational invariance, generated by
$\gamma_5$. One can easily prove that the same holds to all orders
in the gradient expansion. 

 Next, consider the Hubbard-like short-range repulsion term, in the
 continuum notation,
\begin{equation}
 H_U = U \int d^2 \vec{x} ( n_+ (\vec{x}) + n_- (\vec{x}) )^2,  
\end{equation}
with $U>0$. Retaining again only the excitations near the four nodes,
one can write this as
\begin{equation}
S_U = U \int d^2 \vec{r} d\tau [ i \sum_{i=1,2}
\bar{\Psi}_i ' \gamma_5 \gamma_1 \Psi_i ' ]^2. 
\end{equation}
The reader is probably not surprised that
 it is again only $\gamma_5$ that remains
the symmetry generator. This is because $\gamma_5$ in our formalism
is related to translations, which are always the exact
symmetry of the action (2). So both the higher derivative terms, and the
quartic repulsion term reduce
the chiral $SU_c (2)$ subgroup of $U(2)$ to $U_c (1)$, the translations.
One could therefore naively expect that it is the d+ip state, which is
translationally symmetric, that may be preferred by these perturbations.
To decide on this, however, it is not enough
just to know the symmetry of the action, since the new terms $S_1$ and
$S_U$ may turn out to disfavor the d+ip state.
Assuming that both
$\xi ''$ and $\Delta '' $ terms are small, one finds that the contribution
 to the energy of the SDW (or d+ip) state is of second
order in the $S_1$. The interaction term, on the other hand, yields
\begin{equation}
\langle S_U \rangle _0 = -U \sum_i \langle \bar{\Psi}_i ' 
\Psi_i ' \rangle_0 ^2 , 
\end{equation}
with the average taken over the massive $QED_3$ with $\gamma_0 =
\sigma_1 \otimes I $ (cos-SDW). The result, of course,
is the same for the sin-SDW, or for
any linear combination of the cos-SDW and
the sin-SDW. Alternatively, if one assumes
the d+ip ordering, one finds that $S_U$ gives then a {\it positive}
contribution to its energy, 
to the first order in $U$. Although $S_U$ is only translationally symmetric,
it actually {\it inhibits} the formation of the
translationally invariant state, and prefers the ordering to be
in the "orthogonal" direction, i. e. the SDW.

  If both the interaction and the gradient terms are weak, it will
  therefore always
  be the SDW solution that is energetically preferred. This is because
  both the repulsive and the higher derivative terms are equally
  irrelevant by power counting (and have the engineering dimension $-1$),
  the gain in energy due to SDW is of
  first order only in U. The gradient terms affect the energy of the
  SDW only to the next order. So at
  long enough length scales one can alway neglect higher-derivative terms
  as compared to the repulsive interaction, which then serves
  to select the SDW over the d+ip insulator.

\section{Mean-field theory with repulsive interaction}

The message from the previous section is that
the quartic term that represents a short-range repulsion,
although irrelevant, at low but finite energies is
still finite, and it breaks the chiral symmetry
in favor of the SDW state. This is its first important role.
The second is that
once the chiral symmetry is dynamically broken by unbinding
of vortex loops, the quartic term affects the size of the
order parameter, and therefore sets the scale for 
the value of the SDW transition temperature. In this
section I formulate the simplest
 mean-field theory of the chiral symmetry breaking
in presence of the repulsion term, and demonstrate that it  
drastically increases the value of the SDW order parameter at $T=0$. 

We have seen that unbinding of vortex loops leads to   
weak SDW order, but with the order parameter orders of
magnitude smaller than the coupling constant $|\langle \Phi \rangle| ^2 $.
Assuming that the dual condensate as a function of doping
$x$ should be of the same order of
magnitude as the superfluid density on the other side of the
transition (at $x=x_u$),
$|\langle \Phi(x_u -\delta) \rangle| ^2 \sim \rho_{sf} (x_u
+\delta)$, and that Uemura scaling \cite{schneider}
$T_c (x) \propto \rho_{sf}(x) $ is obeyed,
the identification of the size of the SDW order parameter with the
transition  temperature $T_{sdw}(x)$ suggests 
that $T_{sdw}(x_u -\delta) \ll  T_c (x_u +\delta)$.
The difference in the relevant
 scales for the superconducting
and the SDW orderings is in accord with the known phase diagram in the
underdoped regime. Starting from half-filling,
with increased doping the
antiferromagnetic order is quickly lost, and only at a larger
doping the dSC appears. I attribute the absence of the obvious SDW order
very near the superconducting phase to the  inherent weakness of the
spontaneous chiral symmetry breaking in dSC. Assuming that the weak SDW
smoothly evolves into the commensurate antiferromagnet near half-filling,
the obvious problem then becomes the following: how should one 
understand the dramatic increase of $T_{sdw}(x)$ near half filling,
all the way up to $ \sim 300 K$?

\begin{figure}
\centerline{\epsfxsize=7cm \epsfbox{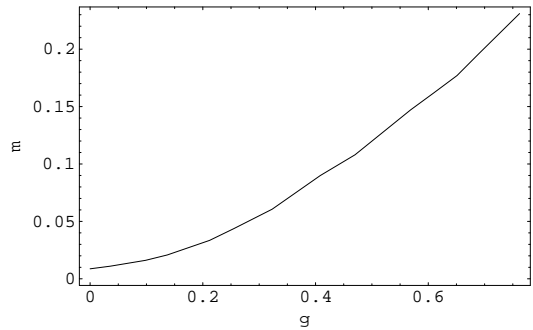}}
\vspace*{1em}

\noindent
\caption{The SDW order parameter $m$ in units of $|\langle \Phi \rangle |^2$
as a function of dimensionless short-range repulsion $g =U|\langle \Phi \rangle
|^2 / (2\pi)^2 $. }
\label{f2}
\end{figure}

 The answer is provided by the observation that although the repulsion
 $U$ is irrelevant
 if weak enough, it enhances the SDW order once it became spontaneously
 generated trough the interaction with the gauge field. To show this I will
 consider the mean-field theory of the $QED_3$ with the additional
 $S_U$  quartic term.
 First, notice that in the Hartree-Fock approximation the $S_U$ term
 gets replaced by the effective quadratic term
 \begin{equation}
 S_U \rightarrow  - U \langle \Psi_i ' \bar{\Psi}_j ' \rangle_0
 \int d^2 \vec{r} d\tau Tr (\bar{ \Psi} _i '
 \gamma_5 \gamma_1 \gamma_5 \gamma_1 \Psi _j ' ) ,
 \end{equation}
  with the average to be calculated self-consistently within the theory
  quadratic in fermionic fields.  The above term corresponds to the
  decoupling in the exchange (Fock) channel, since the direct
  (Hartree) term vanishes. Therefore in the Hartree-Fock approximation,
  after the Franz-Te\v sanovi\' c transformation
\begin{equation}
  S_U \rightarrow U \langle  \Psi _i \bar{\Psi}_i \rangle _0
  \int d^2 \vec{r} d\tau  \bar{\Psi}_i \Psi_i. 
\end{equation}

 Assuming  a uniform 
 $\chi = -U\langle \Psi _i \bar{\Psi}_i \rangle _0 $, and
 treating the gauge-field fluctuations in the large-N approximation
leads to two coupled equations for
$\chi$ and for the momentum dependent fermion self-energy
 \begin{equation}
 \chi= U \int \frac{d^3 \vec{k}}{(2 \pi)^3 }
 \frac{\Sigma(k)}{ k^2 + \Sigma^2 (k) }, 
\end{equation}
\begin{equation}
\Sigma(q) = \chi + 
 \int \frac{d^3 \vec{k}}{(2 \pi)^3 }
 \frac{2 | \langle \Phi \rangle| ^2\Sigma(k)}{ (k^2 + \Sigma^2 (k) ) (  p^2
 +\Pi (p ))  }, 
\end{equation}
with $\vec{p}= \vec{k}-\vec{q}$. 
 When $U=0$ these reduce to the Eq. (32), which leads to $N_c=32/\pi^2$.
 When loops are bound and
 $\langle \Phi \rangle =0$, on the other hand,
 $\Sigma(q) = \chi$, and the Eq. (50) allows a nontrivial solution
 only when the dimensionless coupling
 $g = U \Lambda /(2\pi)^2 > 1$, where $\Lambda$
 is the UV cutoff, $\Lambda < T^*$. Assuming that long-range SDW order
 and dSC
 do not coexist, I take that $g < 1$ in the superconducting phase, so
 that the quartic coupling is there irrelevant.
 With $\langle \Phi \rangle \neq 0$, however, small $g$ ceases to be
 irrelevant, since there is now a small mass scale to 
 effectively cut off its flow.
 Since $\Sigma(q)$ is quickly damped for $q>> |\langle \Phi \rangle| ^2$,
 one can take the UV cutoff in the above equations 
 to be $ \Lambda \sim |\langle \Phi \rangle|^2$. The above equations
 were studied before \cite{carena}, \cite{gusynin} in the context
 of gauged Nambu-Jona Lasinio model of chiral symmetry breaking in
 particle physics. Here I solve the equations numerically
 for $N=2$,  as discussed in the Appendix E. The 
 result is presented in Fig. 4. The main point is that as the
 superconducting phase is more and more disordered and the dual
 condensate grows, the presence of a moderate repulsion  between
 electrons increases the SDW order parameter at $T=0$ by one to two
 orders of magnitude. Recalling the above argument that compares $T_{SDW}$
 to the superconducting $T_c$ on the other side of the
 superconductor-insulator transition, this appears to be 
 in qualitative agreement with the generic behavior observed in
 underdoped cuprates.

\section{Confinement of spinons}

In the superconducting state, the electrically neutral low-energy spinons
represented by the fermionic field $\Psi$ in the $QED_3$ 
are well-defined excitations.
This effective spin-charge separation implicit in
the superconducting state was emphasized in \cite{rokshar},
and more recently in \cite{balents}. One may therefore naturally
wonder if this form of spin-charge separation will survive once
the superconductivity is lost via unbinding of vortex loops.
The answer seems to be {\it no}. It is believed that the chiral symmetry
breaking and confinement go together in the $QED_3$ 
\cite{appelquist}. Qualitative argument why it should be so is provided by the
low momentum form of the polarization tensor in the Eq. (33)
\cite{gopfert}, \cite{burden} : $\Pi (q) \sim q^2 /m $ for $q\ll m$,
so in two dimensions spinons are at large distances bound by a
logarithmic potential. One may independently arrive at the same
conclusion by analytically continuing the fermion propagator in the
broken symmetry phase to real frequencies \cite{maris1},
to find that its poles lie at complex energies with both
real and imaginary parts proportional to the chiral mass.
The chiral symmetry breaking and
confinement of spinons seem therefore to go hand in hand in the $QED_3$,
so the states with broken chiral symmetry,
including most importantly the SDW, should not have well defined
fermionic excitations even above the mass "gap".

Dissapearance of spinons from the spectrum in the insulating phase,
if required, imposes a rather non-trivial constraint on a candidate theory
for underdoped cuprates. For example, one could imagine a completely
different mechanism of chiral symmetry breaking in dSC: even without the 
gauge field, simply increasing the quartic coupling $U$ above a certain value
($U_c \Lambda / (2 \pi)^2 = 1$ in the Hartree-Fock approximation)
would open the gap for spinons and
lead to SDW order. This would be analogous to the chiral
symmetry breaking in the Nambu-Jona Lasinio and related models \cite{nambu},
\cite{rosenstein},
\cite{semenoff}, \cite{renders}. The crucial difference, however, is that
such a mechanism would yield well defined spinon excitations
at energies above the gap, in the insulating state.
The integrity of the gaped
spinons is assured essentially by the Landau phase space arguments.
Such a "deconfined" antiferromagnet was dubbed AF* and studied in
\cite{senthil}, for example.
From this point of view it becomes a non-trivial problem to
understand how spinons could be removed from the spectrum. In the $QED_3$
this is accomplished via the same non-perturbative mechanism that
yields chiral symmetry breaking, described by the Eq. (32), for example.  

Having said all this, it needs to be realized that in a weak SDW
confinement of spinons is effective only over very large distances,
$L>> 1/m$. At intermediate scales, the polarization $\Pi(q)\sim q$, so the
potential between spinons is $\sim 1/r$, and at intermediate distances
$1/m >> L >> 1/|\langle \Phi \rangle|^2 $ spinons will appear
effectively deconfined.
In this sense it is still meaningful to think about underdoped
cuprates as exhibiting an effective spin-charge separation. Computing
the electron spectral function by taking the gauge-field fluctuations
into account in large-N approximation \cite{franz}, which
suppresses the dynamical 
symmetry breaking, for example, gives results in qualitative agreement with
the experiment \cite{rantner}. As one continues to underdope, however,
SDW order parameter grows, and spinons become more strongly confined.
In the strong antiferromagnet at half-filling therefore, on may expect
spinons to be confined already at atomic distances.

\section{Experiment}

  The principal consequence of the $QED_3$ theory of underdoped
cuprates is, of course, the antiferromagnetism itself.  All the
materials that become d-wave superconductors with
doping are insulating antiferromagnets in its parent state.
Furthermore, the sharp spectral features in the dSC should become very broad
in the insulator, since there is a soft (propagator $\sim 1/ q^2$)
gauge field in the problem. Nevertheless, an insulator that derives from a
dSC should partially inherit the d-wave form for its "gap",
except for its finite value in the 
nodal directions. This is in very good agreement with the ARPES
measurement on the insulating $ Ca_2 Cu O_2 Cl_2 $, and
$Sr_2 Cu O_2 Cl_2 $, in its parent state
\cite{ronning}, \cite{ronning1}.
In Fig. 5 I compare the ARPES data for the "gap" measured from the top
of the lower Hubbard band in the insulating state with the simplest
functional form consistent with the chiral mass:
at the remnant Fermi surface 
$\omega= ( (E_{max}(\cos(k_x) -\cos(k_y))/2)^2 + E_{min}^2 )^{1/2}$,
where the chiral mass $m=
E_{min} = 75 meV$ is chosen to be the $T=0$ sublattice
magnetization for $J= 125 meV$. Best fit is obtained
then for $E_{max} = 420 meV$.  The quality of the fit is actually not very
sensitive to some variations in $E_{min}$ and the corresponding $E_{max}$.

\begin{figure}
\centerline{\epsfxsize=7cm \epsfbox{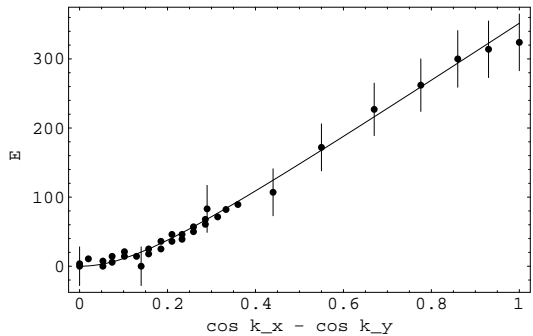}}
\vspace*{1em}

\noindent
\caption{ARPES results for $Ca_2 CuO_2 Cl_2$ (bars) and
$Sr_2 Cu O_2 Cl_2 $ (dots)
with $E = E(k) - E(\pi/2,\pi/2)$ in $meV$. The line is the  function
described in the text. }
\label{f2}
\end{figure}

  The key prediction of this work is that the above "gaped d-wave" form
  of the insulating "gap" is a  generic feature of the insulating state.
  Upon underdoping the ARPES should show the standard d-wave gap for
  sharp quasiparticles in
  the superconducting phase, which should evolve into a gaped
  d-wave form for broad ARPES shape
  in the insulating state, with the "gap" increasing as one
  approaches half filling. The rounding of the data
  at low energies should therefore
  be intrinsic to the insulating state, and should weaken with doping.
  Although the initial experiment on $Ca_2 Cu O_2 Cl_2$
  \cite{ronning} only indicated such
  rounding, later measurements on $Sr_2 Cu O_2 Cl_2 $
  with higher resolution \cite{ronning1}
  clearly showed the deviation from the simple d-wave cusp at lowest energy.
  More recent measurements \cite{ronning2} indicate that the rounding
  of the data at low energies is a robust feature.
  It would clearly be desirable to perform a systematic study of this effect
  at variable doping.

 It may also be worth mentioning that some signs of the
gap rounding in the insulator may be observable already
in the superconducting state.  In $Bi2212$
\cite{mesot}, for example, as one underdopes, the d-wave
gap continues to show the cusp at zero energy, but with the slope
(velocity $v_\Delta$) {\it decreasing}, in spite of the increase in the
overall gap magnitude in $(\pi, \pi)$ direction. It is tempting to 
interpret this effect  as a precursor of the dynamical mass generation.
Detailed study of this effect and of the spectral features in the
insulator is deferred to a future work.

   \section{Conclusion and discussion}

In summary, I have shown that the minimal theory that describes 
unbinding of vortex defects in the d-wave superconductor at $T=0$ is
the two-component, 2+1 dimensional QED,
with the vortex condensate playing the role of
"charge". With the loss of phase coherence, the d-wave
superconductor suffers
the spontaneous breaking of the low-energy "chiral" symmetry,
which results in 
a weak SDW order. It was argued that with underdoping this SDW
smoothly evolves into the strong antiferromagnet near half-filling, with the
selection and the increase of the SDW 
order parameter being provided
by the repulsion between electrons. I argued that spinons are marginally
confined in a weak SDW, and may appear effectively deconfined
over intermediate length scales in the pseudogap regime. Finally,
it was proposed that the rounded d-wave form of the "gap" in the
insulating $Ca_2 CuO_2 Cl_2$ observed by Ronning et al. may be a consequence
of the chiral mass for the approximate
spinon excitations, as  implied by the $QED_3$.

  The present theory is similar in spirit to the approaches of
refs. \cite{balents}, and \cite{senthil}, in that it
attempts to understand the phase diagram of underdoped
high temperature superconductors beginning from the superconducting
phase. It differs, however, in its conclusions to what
the ground state that results from unbinding of topological defects
in the d-wave state is. Whereas it was argued
in \cite{balents} and \cite{senthil} that the relevant description
of this process is provided by the Ising 
$(Z_2)$ gauge theory, and that the resulting state may show spin-charge
separation, I argued that unbinding of defects of unit
vorticity leads to the dynamical symmetry breaking in the
low-energy theory, and the accompanying confinement of spinons
in the insulating state. In fact, if one {\it demands} that the
insulating state near half-filling is the standard antiferromagnet
with spin-1 excitations and confined spinons, the form of a 
single theory that would be able to describe both the dSC and the insulator
becomes severely restricted. The $QED_3$ in this paper is one  such theory.

A variation of the
$QED_3$ as an effective theory for underdoped cuprates has also 
been considered before \cite{affleck}, \cite{marston}, \cite{kim},
\cite{nagaosa} as the theory of low-energy fluctuations around the 
$\pi$-flux phase in the large-N version of the Heisenberg model.
In that approach the gauge invariance reflects the local
particle number conservation at half-filling, and the gauge-field 
has no dynamics on its own. As a result, the gauge field is 
necessarily compact, and the theory is infinitely strongly coupled.
Not much is definitely known about such a lattice gauge theory,
which greatly diminishes its utility. Nevertheless, it was argued
that neglecting  the instanton configurations would restore the
antiferromagnetic order at half filling, via spontaneous 
breaking of a different "chiral" symmetry, which in this
case is actually an enlarged spin rotational symmetry \cite{marston},
\cite{kim}. While this logic may at first 
appear close to the one in the  present work, there are
crucial differences. First, I begin from the superconducting state, away
from half-filling, with the gauge field describing vortex fluctuations. 
As a result, the gauge field is weakly coupled and
non-compact near the dSC-SDW transition.
Also, the SDW phase that obtains from chiral symmetry breaking may be
incommensurate, and the approximate chiral symmetry of the low-energy theory
is unrelated to  spin rotations.

Nevertheless, it may be possible to
understand the $QED_3$ as a  low-energy description of
the microscopic t-J model of cuprates. Starting from the mean-field slave-boson
theory of the t-J model and integrating the constraints of no double
occupancy, for example,  leads to an effective 
theory of the form quite similar to the $QED_3$ \cite{dhlee}, but with the
Volovik's field $\vec{v}$ only. Including vortices would then be expected
to introduce the gauge field $\vec{a}$, as shown in this paper. The
point is that irrespectively from the underlying microscopic model
the theory of the fluctuating dSC should assume the $QED_3$ form.
Values of the parameters, however, may strongly  depend on the
microscopic physics: the bare stiffness $K$ in the XY model
for the phase fluctuations (Eq. (12)), for example, should be
proportional to doping $x$ in the doped Mott insulator \cite{dhlee}.
Also, the charge of quasiparticles (the coefficient in the last term in  the 
Eq. (10))  would be expected to change from unity to $\sim x$, at small
dopings.

 There exist further parallels between the $QED_3$ and the
gauge theory of the t-J model. One may formulate a representation
 of the t-J model with a U(1)
gauge field that minimally couples to spinons and holons. It was argued
\cite{kim} that the effect of holons would be to screen the temporal component
of the gauge field, which then may be shown to halve the critical number
of spinon species for the chiral instability,
$N_c \rightarrow N_c /2$. That way one could avoid the chiral
transition at $N=2$ (assuming that $N_c \approx 3$), and have a spin-liquid
as the ground state in the underdoped regime instead. The tacit
assumption, however, is that uncondensed bosons (holons) at $T=0$ may
exist in a
compressible state.  If the system becomes insulating with the loss
of phase coherence, however, bosons would become incompressible
and the above argument breaks down. This is indeed the case in the
$QED_3$: with the proliferation of vortices the system becomes
insulating,  and all the components of the gauge field become massles.
The same conclusion would be reached within the gauge theory of the t-J
model if one would consider the incompressible state of slave bosons
\cite{ybkim}. 

The present work shares the same
philosophy with the recent one \cite{franz}, \cite{zlatko},
where the massless U(1) gauge
field as an effective description of unbound vortex loops  was also
considered. While the authors \cite{franz} considered the large-N limit
of the $QED_3$, and thus
precluded the chiral symmetry breaking, my main point is that at $T=0$
the spontaneous formation of the chiral condensate is nothing else
but the SDW instability of the d-wave superconductor.
The results of the ref. \cite{franz} may therefore be understood
as applying to the finite-T phase much below the pseudogap scale $T^*$ in Fig. 1.

The problem of phase disordering of dSC has also been recently studied
by Ye \cite{ye}. Working in the Anderson gauge  \cite{anders}
in which $\phi_{sA} = \phi_s$, $\phi_{sB} = 0$ in the Eq. (8),
the author concluded that the gauge-field $\vec{a}$ is
always massive when charge fluctuations are included. It is easy to
see that this is a direct consequence of the gauge choice: in the Anderson
gauge $\vec{a}= \vec{v}$, and not only $\vec{v}$, but $\vec{a}$ too is 
ultimately coupled to the charge current. In my gauge invariant approach,
on the other hand, $\vec{a}$ is completely
decoupled from charge, and couples only to spin. Inclusion of charge
fluctuations therefore does not make $\vec{a}$ massive, but simply
adds an irrelevant quartic coupling to the $QED_3$ Lagrangian. 

 The intimate relationship between the d-wave superconductivity and
 antiferromagnetism is also the main theme of the $SO(5)$
 theory of Zhang \cite{zhang}. The present work echoes some of that
general idea, but is based on entirely different physical principles. 
In particular, although there should be 
 a direct dSC-SDW transition in the phase diagram, this appears  
 unrelated to the $SO(5)$ symmetry, but comes as a consequence of the
{\it chiral} symmetry that emerges at low energies in the d-wave superconducting state.
It is the spontaneous breaking of this hidden approximate symmetry that 
implies then the breaking of the spin rotational symmetry in the SDW phase.

 Marginal confinement of spinons we found in the weak SDW phase is very much
in line with the speculations of Laughlin \cite{laughlin}, \cite{fradkin} on 
parallels between the antiferromagnetism and the
confinement in strong interactions. In fact, the 
$QED_3$ shows precisely how chiral symmetry breaking, i. e. SDW
ordering, binds spinons into spin-1 objects. 
Deconfinement in this theory seems indeed
tantamount to the absence of chiral symmetry breaking.
In this context, it may be interesting to note 
that the $d+id$ state, that would correspond to
the $i\tilde{\gamma}_1 \tilde{\gamma}_2 $ matrix in (38), could
lead to deconfined spinons. This state is outside of the chiral manifold,
and it is believed that it is not
spontaneously induced in the $QED_3$ \cite{appelquist}, because of the
Chern-Simons term that becomes generated for the gauge-field.
With the Chern-Simons term, on the other hand, the gauge-field propagator
behaves like $\sim q$ at low momenta, and thus spinons may become
deconfined \cite{shaposhnik}. The chiral symmetry breaking in the $QED_3$ is therefore nothing
by the effective description of the spinon confinement.

  It is also interesting to note that were the critical number of
fermions $N_c<2$, the result of phase disordering of dSC would be
quite different. Instead of symmetry breaking and confinement one would
find a gapless,
chirally symmetric state, in which spinons would be deconfined. This is
again because the polarization tensor would then be $\sim q$ at low momenta,
i. e. the interaction between spinons would be
$\sim 1/r$ at large distances. This state would be similar in spirit to the 
"nodal liquid" \cite{balents}, or
analogous to the "algebraic Fermi liquids"
\cite{franz}, \cite{rantner},\cite{wen1}
proposed in literature as candidates for
the pseudogap phase.  It has been proposed recently that $N_c = 3/2$
exactly \cite{cohen}, although all the actual calculations based on
Schwinger-Dyson formalism lead to $N_c >3$. If $N_c$ is indeed that low,
phase disordering of the dSC
would first lead to the deconfined pseudogap phase,
which only later would turn into the confined SDW phase, presumably
due to the repulsive quartic term which is know to 
increase $N_c$ \cite{carena}, \cite{gusynin}.
At this time it is hard to say which
one of these two scenarios is realized in cuprates.

The main point made in this paper is that unbinding of
vortex loops in a d-wave superconductor at $T=0$ 
results in SDW order. It then appears natural to assume
that the cores of fluctuating vortices are
already in the insulating state.
This speculation is in accord with the recent STM, neutron
scattering, and NMR experiments \cite{renner}, \cite{hoffman}, \cite{lake},
\cite{vesna}, 
the SO(5) proposal \cite{zhang}, \cite{arovas}, and the mean-field
\cite{zhu} and the finite size $QED_3$ calculations \cite{sheehy}.
The superconductor-insulator transition would then be the result of
the decrease of the bare stiffness $K$ in the XY model
with underdoping, since $K\sim x$ in the doped Mott insulator \cite{dhlee}.

\section{Further problems}

  I finish with a tentative list of problems opened by this
work.

1) The role of strong anisotropy $v_f /v_{\Delta} \gg 1 $
that exists in cuprates is unclear. In particular, since anisotropy
on the bare level is marginal, it may affect the value of $N_c$. The
preliminary results, indicate, however, that weak anisotropy is
irrelevant, so that one would expect
$N_c$ to be unaffected by it \cite{dom}.

2) Nature of the various phase transitions in the theory
is also of interest. Whereas one expects that 
gapless quasiparticles do not change the Kosterlitz-Thouless
universality class of the finite temperature superconducting transition, the
nature of chiral symmetry breaking at finite temperature
and its possible interplay with Neel transition is far less clear
\cite{aitchison}. In particular, in relation to the Uemura scaling
\cite{schneider}, one would like to understand the quantum
superconductor-insulator criticality and how it may be affected
by gapless spinons.

3) Can long-range SDW and SC order coexist? In the approximation
employed in the present work, the gauge-field $\vec{a}$ is considered
decoupled from spinons in the dSC phase. This is likely to underestimate
the effect of $\vec{a}$, and a better approximation for the gauge-field
propagator is needed to study its effect {\it inside} the dSC. This could be
important in light of the recent experimental data \cite{lake},
\cite{miller} that may be interpreted as indicating the
coexistence of the SDW and the SC orders in some compounds \cite{demler}.

4) The present work also points to a new route
towards a deconfined phase in two dimensions: lowering $N_c$ below
two would allow for an insulating phase with deconfined spinons. At
present, however, it is not clear how to achieve this within the
$QED_3$, unless the Schwinger-Dyson equations systematically
overestimate $N_c$ \cite{cohen}.

5) The computation of the electron propagator within the
$QED_3$ is an important problem \cite{khvesch}. This would be 
necessary for a detailed comparison of the theory with the
ARPES measurements.

6) As mentioned at the end of sec. IV, in the Villain approximation
to the XY model,  the gauge-field
$\vec{a}$ appears to be compact, in contrast to the Volovik's field $\vec{v}$.
Although this should be an artifact of the Villain approximation, it
would still be interesting to understand the 
effect of compactness of $\vec{a}$ on the chiral symmetry
breaking in the $QED_3$. It has been
argued that the coupling to gapless spinons makes the single instanton
anti-instanton pair that derives from compactness of $\vec{a}$
bound above the certain number of spinon components $N_{inst}$ \cite{marston},
\cite{kleinert}. $N_{inst}$ may be made smaller than
$N_c$ for chiral symmetry breaking by a large anisotropy \cite{wen1},
for example. It is unclear, however, whether this conclusion survives the effects
of screening by other pairs \cite{subir}. Also, even
if the instantons can be made
 irrelevant above $N_c$, below $N_c$ one would expect
them to become relevant again with the opening of the spinon "gap".
This in turn could have profound consequences
for the spinon confinement.

 \section{Acknowledgement}

 The author has benefited greatly from discussions and exchanges
 with S. Dodge, V. Gusynin, M. P. A.  Fisher,  E. Fradkin, M. Franz, K. Kavehmaryan
 D. J. Lee, B. Marston, M. Reenders, F. Ronning, S. Sachdev,
 B. Seradjeh, S. Sondhi, Z. Te\v sanovi\' c,  C. Wu, W.-C. Wu,
 and S.-C Zhang. This work was supported
 by NSERC of Canada and the Research Corporation.

 \section{Appendix A}

I present the self-consistent mean-field theory
of the lattice superconductor (20) \cite{peskin}, and use it to
approximately compute the correlator appearing in the Eq. (20).
By the Bogoliubov inequality: 
\begin{equation}
Z_{xy} \geq Z_0 e^{- \langle H-H_0 \rangle_0 },
\end{equation}
where $Z_{xy}$ is the partition function in the dual form (21) with a 
finite "inverse temperature" $x$,
and the average in the exponent is performed
over a {\it local} mean-field Hamiltonian
\begin{equation}
H_0 = -h \sum \cos \theta_i + \frac{1}{8K \pi^2}\sum(\nabla \times \vec{\Phi})^2
+ \frac{m^2}{4 K \pi^2} \sum \vec{\Phi}^2. 
\end{equation}
The optimal values of the parameters $h$ and $m$ that maximize the right hand side
in the Bogoliubov inequality are then determined by the equations:
\begin{equation}
h= \frac{6A}{x} \frac{I_1 (h)}{I_0(h)}, 
\end{equation}
\begin{equation}
m^2 = \frac{K \pi^2}{3} \frac{I_1 (h)}{I_0(h)} h , 
\end{equation}
\begin{equation}
A= exp [ -\frac{2 K \pi^2}{3} \int \frac{d^3 \vec{k}}{(2\pi)^3} 
\frac{1}{ F(k) +m^2 } ], 
\end{equation}
where $F(k) = \sum_\nu (e^{i k_\nu} -1)^2$, and the integral over
$\vec{k}$ is taken over $(-\pi,\pi)$. $I_0$ and $I_1$ are the
Bessel functions. These equations can be solved
graphically, and describe a discontinuous transition from the
phase with $h=m=0$ (bound vortex loops), to the condensed
phase $h\neq 0, m\neq 0$ (infinitely large vortex loops) \cite{balian}.

  The requisite average in the Eq. (20)
is easy to compute in the mean-field theory that has different
sites decoupled:
\begin{equation}
\langle cos( \theta_i - \theta_{i+\hat{nu} } - \Phi_{i, \nu}) \rangle_0
= |\langle e^{i \theta_i} \rangle_0 |^2 
\langle e^{-i \Phi_{i, \nu}}\rangle_0. 
\end{equation}
Since, $\langle e^{-i \Phi_{i, \nu}}\rangle_0 = A$
and finite, we conclude that
 \begin{equation}
\langle cos( \theta_i - \theta_{i+\nu} - 2 \pi \Phi_{i, \nu}) \rangle_0
\propto h^2 , 
\end{equation}
i. e. finite {\it only} in the ordered phase of the dual theory (20),
i. e. in  the disordered phase of the original XY model.

\section{Appendix B}

Here I provide a different
derivation of the dynamics of the gauge-field $\vec{a}$
at $T\neq 0$ starting from the Hamiltonian for the Coulomb plasma.
Assume a collection
of $N_+ (N_-)$ vortices (antivortices) at the positions
$\{ \vec{r}_i \}$. The Hamiltonian of the vortex system is
\begin{equation}
H_v = \frac{1}{2} \sum_{i=1}^N  q_i q_j v(\vec{r}_i - \vec{r}_j), 
\end{equation}
where $v(\vec{r})\approx   -\ln |\vec{r}|$, at large distances,
and $N=N^+ + N^- $, $q_i = \pm 1$.  The partition function
of the vortex system $Z_v$ can then be written as
\begin{eqnarray}
Z_v = \sum_{N_{A,B}^{+,-}=0}^{\infty} 
\frac{N^+ !}{ N_A ^+ ! N_B^+ !}
\frac{N^- !}{ N_A ^- ! N_B^- !}   \\ \nonumber
\frac{(y/2)^N }{ N^+ ! N^- !}
\int \prod_{i=1} ^{N} d\vec{r}_i
e^{-\frac{H_v}{T}} , 
\end{eqnarray}
where $N^{+(-)} = N_A^{+(-)} + N_B^{+(-)}$, and $y$ is the bare
vortex fugacity. The combinatorial factors serve to ensure that
in  $Z_v$ one sums over {\it all} possible divisions of vortices
and antivortices
into groups A and B, and divides by the number of combinations.
With this symmetrization the symmetry between up and down spin
in the original Hamiltonian (2) will be preserved in the
Dirac theory for neutral spinons. 
This also guarantees that on average there is an equal number
of vortices (and antivortices) in both groups.

Next, introduce the vorticity 
densities in $Z_v$ by inserting the unity 
\begin{equation}
1= \int D[\rho_{A}] \delta( 
\rho_{A} (\vec{r}) - \sum_{i=1}^{N_{A}} q_{iA}\delta ( \vec{r}
- \vec{r}_{iA})),   
\end{equation}
 and similarly for B. The gauge field then  becomes
\begin{equation}
(\nabla\times \vec{a}(\vec{r}) )_{\tau}  = \pi (\rho_A (\vec{r}) 
- \rho_B(\vec{r}) ), 
\end{equation}
in the transverse gauge $\nabla\cdot\vec{a}=0$, and the index denotes the
$\tau$ component.
$\vec{v}$ is defined the same way except with the plus sign between
$\rho_A$ and $\rho_B$.

 By introducing two auxiliary fields $\Phi_A$ and
 $\Phi_B$ to enforce the constraints, after the 
 integration over the densities the partition function may be written as 
\begin{eqnarray}
Z_v = \sum_{N_{A,B} ^{+,-}=0}^{\infty}
\frac{(y/2) ^N }{N_A^+ ! N_A^- ! N_B^+ !  N_B^- !}
\int D[\vec{a}, \vec{v}, \Phi_+, \Phi_-] \\ \nonumber
\exp -[ \frac{1}{2 \pi^2 T} \int d\vec{r} d\vec{r}' B(\vec{r}) v(\vec{r}
-\vec{r}') B(\vec{r}')  \\ \nonumber
+\frac{i}{2\pi} \int d\vec{r} (B(\vec{r}) \Phi_+(\vec{r})+   
 b(r) \Phi_-(\vec{r}) )   \\ \nonumber
 - \sum_{i\alpha, \alpha=A,B}
\ln \int d\vec{r} \exp(i q_{i\alpha} \Phi_\alpha (\vec{r}) ) ] , 
\end{eqnarray}
where $\Phi_{+,-} = \Phi_A \pm \Phi_B$, $B(\vec{r})=
(\nabla\times \vec{v})_\tau $,
and $b(\vec{r})= (\nabla\times \vec{a})_\tau$.
Performing the summations yields
\begin{eqnarray}
Z_v= \int D[\vec{a}, \vec{v}, \Phi_+, \Phi_-] \\ \nonumber
\exp -[ \frac{1}{2 \pi^2 T} \int d\vec{r} d\vec{r}' B(\vec{r}) v(\vec{r}
-\vec{r}') B(\vec{r}') \\ \nonumber 
+\frac{i}{2 \pi} \int d\vec{r} (B(\vec{r}) \Phi_+(\vec{r})  +
 b(\vec{r}) \Phi_-(\vec{r})) \\ \nonumber
 - y \int d\vec{r} ( \cos \Phi_A(\vec{r})
 + \cos \Phi_B(\vec{r}) ) ].
\end{eqnarray}

 Finally, neglecting the coupling to the charge current,
 the Gaussian integration over $\vec{v}$ (i. e.  $B$) gives
\begin{eqnarray}
Z_v= \int D[\vec{a},\Phi_+, \Phi_-] 
\exp -[ \int d\vec{r} [ \frac{T}{2}  (\nabla \Phi_+)^2  + \\ \nonumber 
+\frac{i}{\pi}  b(\vec{r}) \Phi_-(\vec{r})  
- 2 y \cos (\Phi_+(\vec{r}))  \cos (\Phi_-(\vec{r}) ) ], 
\end{eqnarray}
where I also have rescaled the $\Phi$ fields by a factor of two.
The last expression is then analogous to the $T=0$ expression in
the Eq. (18) with $x$ finite and without the dual angles $\theta_{A,B}$.
By introducing a source term in the action, $\sim  i\int  j(r) b(r) /
\pi$, and integrating
over $b$, one readily finds  
\begin{equation}
\langle ( \nabla\times\vec{a}(\vec{r}))_{\tau}
(\nabla\times\vec{a}(\vec{r}'))_{\tau} 
\rangle = \langle y  \rangle
\delta(\vec{r}-\vec{r}'),  
\end{equation}
where $\langle y \rangle  = y \pi ^2
 \langle \exp ( i \Phi_+) \rangle$, with the
average to be taken at $\Phi_- = \vec{a}\equiv 0$.
One recognizes $\langle y\rangle$ as the renormalized, or running,
fugacity in the Kosterlitz-Thouless scaling, which signals
the appearance of free vortices. $\langle y \rangle $ plays the
role analogous to the vortex loop condensate in 2+1 dimensions, in
providing a mass for the field $\Phi_+$ in the Eq. (65). 
This implies the Maxwell term at $T\neq 0$
for the $\tau$ component of $\nabla \times
\vec{a} $ once fluctuating vortices are integrated out. 

 \section{Appendix C}

For completeness, here I outline the derivation of the
result that chiral symmetry in isotropic massless QED3
is spontaneously broken for $N<N_c$, with $N_c$ finite,
at any value of the coupling constant.

  Rescaling the momenta $p/m \rightarrow p$ and self-energies
  $\Sigma(p) /m \rightarrow m$ and $\Pi(p) /m^2 \rightarrow \Pi(p)$,
  after taking the limit $q \rightarrow 0$ in the Eq. (32) we find
  \begin{equation}
 1= \frac{|\langle \Phi \rangle |^2} {\pi^2 m} \int_0 ^{\Lambda/m} dp
 \frac{p^2 \Sigma(p)}{(p^2 + \Sigma^2 (p)) ( p^2 + \Pi(p) ) }, 
 \end{equation}
 where  the polarization is now
 \begin{equation}
 \Pi (p) = \frac{N |\langle \Phi\rangle|^2 }{ 4 \pi m}  f(p),
 \end{equation}
 with 
 \begin{equation}
 f(p)= 
 ( 2+ \frac{p^2 -4}{p} \sin^{-1} ( \frac{p}{\sqrt{ 4 + p^2} } ), 
 \end{equation}
to the leading order in $N$ \cite{appelquist}.
We see that the right-hand side of the
Eq. (67) is a decreasing function of $m$, so for $m\neq 0$ solution to
exist we just need $RHS(m=0) >1$. This is satisfied for $N<N_c$,
where
\begin{equation}
N_c = 4 \int_0 ^{\infty} dp
\frac{p^ 2 \Sigma(p)}{( p^2 + \Sigma^2 (p) ) f(p)}. 
\end{equation}
As defined, $\Sigma(0)=1$, and one expects $\Sigma(p)$ to vanish at large
momenta. 
Also, $f(p)\approx \pi p /2$ for $p>>1$, so the integrand at large
argument behaves like $\sim \Sigma(p)/p $. $N_c$ is therefore finite,
and independent of the coupling constant $\langle \Phi \rangle$.
Its precise value in the
large-N approximation will depend only  on the
function $\Sigma(p)$ at $N=N_c$, and can be obtained by solving
the differential equation equivalent to the integral equation (67)
\cite{appelquist} (see Appendix E). This yields $N_c = 32/\pi^2$, not
far from the results of other more elaborate computations that go beyond
the leading order in $N$ \cite{maris}, \cite{kocic}.

\section{Appendix D}

Here I discuss a different representation of the quasiparticle
action, more in line
with the previous work \cite{balents}. This should serve to underline 
the difference between the approximate chiral $SU_c (2)$ symmetry, and the
exact spin rotational $SO(3)$, also present in dSC. It is only
the latter that will appear in the different version of the theory 
considered here and in \cite{balents},
while the chiral symmetry will remain completely obscured.

  I start again from the same quasiparticle action in the Eq. (2),
but now introduce the four-component field as
\begin{eqnarray}
\Psi^{' \dagger}_{1(2)} (\vec{q},\omega_n ) = (c^{\dagger}_+ (\vec{k},\omega_n ),
c_- (-\vec{k}, -\omega_n ), \\ \nonumber
c^{\dagger}_- (\vec{k}, \omega_n ),
- c_+ (-\vec{k}, -\omega_n ) ). 
\end{eqnarray}
By linearizing the spectrum and by retaining only the modes near the four
nodes, the continuum theory may again  be written as 
\begin{eqnarray}
S[\Psi'] = \int d^2 \vec{r} \int_0 ^ {\beta} d\tau \Psi_1 ^{' \dagger} [
\partial_\tau
+ M_1 v_f \partial_x + M_2 v_{\Delta} \partial_y] \Psi_1 '\\ \nonumber
 + ( 1 \rightarrow 2, x\leftrightarrow  y) , 
\end{eqnarray}
 but this time with different form of the matrices $M_1$ and $M_2$: 
 $M_1= -i I \otimes \sigma_3 $, and $M_2 = i I \otimes \sigma_1 $.
 Introducing $\gamma_0 = \sigma_3 \otimes \sigma_2$, for example,
 the theory becomes
\begin{eqnarray}
S[\Psi' ] = \int d^2 \vec{r} \int_0 ^ {\beta} d\tau \bar{\Psi} _1 ' [
\gamma_0 \partial_\tau
+ \gamma_1 v_f \partial_x + \gamma_2 v_{\Delta} \partial_y] \Psi_1 '
\\ \nonumber
 + ( 1 \rightarrow 2, x\leftrightarrow y ), 
\end{eqnarray}
with $\gamma_1 = \sigma_3 \otimes \sigma_1$ and $\gamma_2=\sigma_3
\otimes \sigma_3 $. It is interesting to consider the generators
of the global $U(2)=U(1)\times SU(2)$ symmetry per Dirac
component present in this representation of the theory.
They are $I_4= I \otimes I $, $\gamma_3 = \sigma_1 \otimes I$,
$\gamma_5 = -\sigma_2 \otimes I$, and $\gamma_{35}=\sigma_3 \otimes
I $, respectively.
One may recognize the $U(1)$ factor as representing now the continuous
translations, since under a translation 
$ c_\sigma (\vec{k}, \omega) \rightarrow e^{i\vec{k}\cdot \vec{R}}
c_\sigma (\vec{k}, \omega)$, the Dirac field  now transforms as
\begin{equation}
\Psi_i ' (\vec{r},\tau) \rightarrow e^{ i \vec{K}_i \cdot
\vec{R}} \Psi_i ' (\vec{r}+\vec{R},\tau). 
\end{equation}
The $SU(2)$ operators, on the other hand, are nothing but the {\it spin
rotations}. In fact, the above $U(2)$ is an exact
symmetry of the Hamiltonian (2), and is present even if all higher
order derivatives are retained.

  Including the coupling to vortex loops via massless gauge field
  in the above representation of the problem then
  may spontaneously induce only the d+ip insulator. This breaks two of the
  above generators, which then simply rotate the spin-axis.
  Translational symetry is, on the other hand, always preserved in this
  formulation, and the SDW remains invisible.

\section{Appendix E}

Here I provide the details behind the numerical solution of the
Eqs. 50-51. Since we are interested only in the qualitative effect
of the U-term, it will suffice to assume that the fermion mass
is small, $m \ll |\langle \Phi \rangle| ^2 $, so that one can neglect
the $p^2$ term compared to  $\Pi(p)$ in the Eq. (51), and take
\begin{equation}
\Pi(p) = \frac{N |\langle \Phi \rangle| ^2 }{ 8} p , 
\end{equation}
appropriate for $p >> m$. This approximation is known to lead to
even quantitatively good result for the mass
for $N$ as low as unity, when $U=0$ \cite{appelquist}.
Performing the angular integrals then gives
\begin{equation}
\Sigma(q) = \chi + \frac{8}{N \pi^2 q } \int_0 ^{\Lambda} dk 
\frac{k \Sigma(k) ( k - (k-q) \theta(k-q) ) }{k^2 + \Sigma^2 (k)}. 
\end{equation}
Differentiating twice one finds that this integral equation
is equivalent to the differential equation \cite{appelquist}:
\begin{equation}
\frac{d}{dq} (q^2 \frac{d}{dq} \Sigma(q)) = -\frac{8}{N \pi^2}
\frac{q^2 \Sigma(q) }{ q^2 + \Sigma^2 (q) },
\end{equation}
with the boundary condition
\begin{equation}
\Lambda \Sigma' (\Lambda) + \Sigma(\Lambda) = \chi, 
\end{equation}
and with 
 \begin{equation}
 \chi = \frac{U}{(2\pi)^2} \int_0 ^{\Lambda} dq \frac{q^2 \Sigma(q)}
 { q^2 + \Sigma^2 (q)}. 
 \end{equation}
Here I take $\Lambda = |\langle \Phi \rangle |^2 $.

 The above equations may now be studied by assuming
 $q \gg \Sigma(q)$, which leads to linear equation that can be
 exactly solved \cite{miransky}. This yields, for example,
 the well known transition line in the $g-N$ plane:
 $ g_c (N) = (1/4) (1+ \sqrt{1- (N_c/N)^2}$, for $ N>N_c $,
 $g_c \leq 1/4$  for $N=N_c$, 
 with $N_c = 32/\pi^2 $. To determine the size of
 $\Sigma(0)$, however, one needs to solve the full non-linear
 equation. This may be accomplished, for example,
  by choosing a value for $\chi$,
assuming $\Sigma(\Lambda)$ next, and then iterating back to find
$\Sigma(q)$ for $0<q<\Lambda$. The solution is found by tuning
$\Sigma(\Lambda)$ to achieve $\Sigma(0)$ finite.
One then computes the value of $g = U \Lambda /(2\pi)^2$
from the assumed
$\chi$ and the found $\Sigma(q)$. This procedure leads to the Fig. 4.


\begin{thebibliography}{99}
\bibitem{rem} Definition of $\vec{v}$ here differs
slightly from the one below
the Eq. (10), in that it does not contain the spin-wave contribution
which has already been integrated out in arriving at the Eq. (13).
\end{thebibliography}
\end{document}